\documentclass{aa}

\usepackage{graphicx}
\usepackage{txfonts}
\usepackage{lipsum}
\usepackage{multirow}
\usepackage{xcolor}
\usepackage{hyperref}
\usepackage{amsmath,siunitx}
\DeclareUnicodeCharacter{2212}{-}
\usepackage{textcomp}
\usepackage{lineno}
\usepackage{linenoaa}

\begin{document}

   \title{Magnetic field, rotation, and binarity of the first magnetic B[e] star, IRAS 17449+2320}

   \author{I. Bermejo Lozano
          \inst{1}
          \and
          G. A. Wade\inst{2,3}
          \and
          C. P. Folsom\inst{4}
          \and
          D. Kor\v{c}\'{a}kov\'{a}\inst{1}
          }

   \institute{Charles University, Faculty of Mathematics and Physics, Astronomical Institute, V Hole\v{s}ovi\v{c}k\'ach 2, 180 00 Praha 8, Czech Republic\\
              \email{irisberloz@gmail.com}
         \and
             Department of Physics, Engineering Physics \& Astronomy, Queen’s University, Kingston, Ontario K7L 3N6, Canada
        \and
            Department of Physics \& Space Science, Royal Military College of Canada, PO Box 17000, Station Forces, Kingston, ON K7K 7B4, Canada
        \and
            Tartu Observatory, University of Tartu, Observatooriumi 1, Tõravere, 61602, Estonia
             }

   \date{\today}

  \abstract
   {IRAS 17449+2320 is the first identified magnetic B[e] star. It belongs to a subgroup of peculiar hot B-type stars, called FS~CMa stars. All these objects present the B[e] phenomenon, i.e., permitted and forbidden emission lines, and infrared excess as a result of the large amount of gas and dust surrounding the central star. The origin of this circumstellar material is unclear. Most studies suggest that it is the result of a mass flow or transfer in a close binary system. However, the circumstellar matter and the presence of the magnetic field in IRAS 17449+2320 point to the possibility that this star is the result of a merger.}
   {In this paper we present the recent work done to confirm and characterize the magnetic field of this object, and constrain its rotation and binarity.}
   {The sample used for this work consists of 12 spectropolarimetric observations taken with ESPaDOnS at the Canada-France-Hawaii Telescope, obtained from 2006 to 2025. The circular polarization (Stokes $V$) was measured in each observation, and linear polarization (Stokes $Q$ and $U$) was measured in two of them. We obtained the least squares deconvolution (LSD) profile of the Stokes $V$ parameter of each spectropolarimetric observation. From the LSD profiles, we measured the longitudinal magnetic field (B$_z$), and obtained the magnetic field modulus ($|B|$) from the Zeeman splitting of some metallic lines of the spectra. We calculated a Lomb-Scargle periodogram over the B$_z$ dataset to obtain the rotational period of this star. Using the unpolarized (Stokes $I$) spectrum, we performed precise measurements of the equivalent widths and radial velocities of various spectral lines.}
   {The longitudinal magnetic field presents a sinusoidal variation with a stable period of 36.11 days, with values ranging from 3\,000 G to -1\,000~G. The magnetic field modulus measurements depend somewhat on the metallic line used to measure it. For \ion{N}{I}, the values range from $\sim$4\,600~G to $\sim$5\,500~G; from $\sim$4\,400~G to $\sim$5\,500~G for \ion{Si}{II}, and from $\sim$4\,300~G to $\sim$6\,000~G for \ion{Ca}{II}. The magnetic field modulus shows a periodic variation for the three different metallic lines, although they require a second order term to perform a better fit on the datasets. The stable sinusoidal variation of the longitudinal field implies an approximately dipolar surface field topology. Some lines of the spectra show an equivalent width variation with time, following the period of the star. We also obtained the radial velocity from each LSD Stokes I profile, and calculated a mean value of -17.6 km s$^{-1}$.}
   {IRAS~17449+2320 is the first magnetic B[e] star detected. It presents Zeeman splitting in individual lines and consistent Stokes $V$ profiles. We see no evidence of binarity in this object, with stable radial velocities and no detectable lines of a secondary star in the spectrum. We consider the possibility that this star is the result of a merger, which could have generated the strong magnetic field and the large amount of circumstellar material.}


   \maketitle
%
\nolinenumbers
\section{Introduction}\label{sec:Introduction}
The first B[e] stars were introduced by \cite{Allen1976}. They discovered that some B-type stars with optical emission lines in our galaxy also showed forbidden emission lines of low ionization and infrared excess, which distinguished them from classical Be stars \citep{Rivinius2013}. However, there were many different types of objects with these characteristics. \cite{Lamers1998} performed the first classification of them based on their gas properties and proposed changing the concept of B[e] stars to the B[e] phenomenon. They identified four main groups of such objects: compact planetary nebulae, symbiotic stars, Herbig Ae/Be stars, and B[e] supergiants. However, there were around 50 objects that could not be included in any of these subgroups, the unclassified B[e] stars.\smallskip

Later, \cite{Miroshnichenko2007} called almost all of this unclassified group FS~CMa stars and defined them as a subgroup of peculiar hot stars, mainly B-type, but with a wide range of spectral types, from O9 to A2. They are characterized by the B[e] phenomenon, this means that, specifically for FS CMa stars, they show forbidden emission lines of low-ionization metal lines, such as \ion{O}{[I]} or \ion{Fe}{[II]}, as well as permitted emission lines. The Balmer line emission is typically unusually strong. They also present an infrared excess around 20 $\mu$m that sharply decreases toward longer wavelengths, and all of them are located far from star formation regions. There are around 50 confirmed members in this subgroup and around ten additional candidates\footnote{The publicly available list of FS~CMa stars can be found at \href{https://home.uncg.edu/~a_mirosh/hswd/main/FSCMa_objects_full_list_Sept2013.html}{https://home.uncg.edu/$\sim$a\_mirosh/hswd/main/FSCMa\_objects\_full\_list\_\\ Sept2013.html}}. However, in a recent publication, \cite{Miroshnichenko2023} increased the number to 70, but without providing a list of the members.\smallskip

It is broadly agreed that these characteristics are produced by large amounts of gas and dust surrounding the central star, the origin of which is still unclear. The main scenario accepted to date is a binary origin \citep{Miroshnichenko2002, Mennickent2017, Khokhlov2018}. Two different binary configurations have been suggested in order to explain the characteristics of these stars: a binary system in a close orbit with a recent mass outflow phase due to Roche Lobe overflow (RLOF)  \citep{Miroshnichenko2007, Miroshnichenko2011}, or a binary system with one intermediate-mass post-Asymptotic Giant Branch (AGB) component at the beginning of the ejection of a planetary nebula with a main sequence companion \citep{Miroshnichenko2006, Miroshnichenko2013}.\smallskip

There are some observational arguments used to support the binary scenario: the large inferred mass-loss rate (\.M), regular periodicity, composite spectra, and the presence of lithium lines in some cases. The \.M has been calculated for three different FS~CMa stars, HD~87643 (\.M = 7 $\times$ 10$^{-7}$ M$_{\odot}$ yr$^{−1}$; \citealt{FreitasPacheco1982}), AS~78 (\.M = 1.5 $\times$ 10$^{-6}$ M$_{\odot}$ yr$^{-1}$ ; \citealt{Miroshnichenko2000a}), and IRAS~00470+6429 (\.M = (2.5 - 2.9)$\times$10$^{-7}$ M$_{\odot}$ yr$^{-1}$ ; \citealt{Carciofi2010}). The methods used to estimate these \.M values were the comparison with the \ion{Fe}{II} multiplet 42 lines, the spherical model, and the Pogodin (1986) code to calculate the mass-loss rate by fitting H$\beta$, and using a non-Local Thermodynamic Equilibrium (LTE) Monte Carlo radiative transfer code (HDUST), respectively. These values of mass-loss rates cannot be produced by a single non-supergiant B-type star. However, the \.M for these stars was calculated using the classical Castor-Abbott-Klein (CAK) wind model. Further observations have shown that the circumstellar matter is accumulated closer around the star, and therefore the theoretical models give an overestimated value of the mass-loss rate \citep{Kucerova2013}.\smallskip

There are two FS~CMa stars that present a regular variability in their radial velocity, MWC~728 \citep{Miroshnichenko2015} and AS~386 \citep{Khokhlov2018}, and two more that present a composite spectrum (i.e., with sets of lines from two different stars), MWC~623 \citep{Zickgraf2001} and V~669~Cep \citep{Miroshnichenko2002}. For MWC~623, it is considered that the binary system consists of a B4 III star with a K2 II-Ib star as the companion. However, it is not clear that this composite spectrum belongs to a binary system. The variability of the H$_{\alpha}$ bisector could not be properly modeled by a secondary for MWC~623 \citep{Polster2018}, and polarimetric observations suggest that some lines could have been formed in the pseudo-atmosphere of the disk \citep{Zickgraf1989}. V~669~Cep was classified as a binary system by comparing its spectrum to that of MWC~623, as both of them show similar spectroscopic characteristics.\smallskip

Additionally, some FS~CMa stars present the lithium doublet at 6708\AA~ in their spectra. \cite{Korcakova2020} looked for these lithium lines in 16 FS~CMa stars and found this doublet in eight of them. Lithium lines are incompatible with a B-type star, therefore, these lines are considered to be part of a cool companion, i.e., an unresolved binary system. However, these lithium lines are not necessarily photospheric. They may be circumstellar, and thus a secondary object is not required to explain the presence of lithium lines from some of these stars \citep{Alecian2013, Korcakova2025}.\smallskip

\cite{Miroshnichenko2023} report 21 FS~CMa stars with some evidence of a binary, but in most of these cases the evidence is indirect or circumstantial. Moreover, there are some recent studies that have discarded objects from the FS~CMa group. \cite{Aret2020} recently cataloged CI~Cam as a supergiant B[e] high-mass X-ray binary, and AS~386 is suggested to be a binary system with a black hole as one of the components \citep{Khokhlov2018}.\smallskip

\cite{delaFuente2015} introduced a new hypothesis to explain the origin of some FS CMa stars after discovering two FS~CMa stars in young massive clusters: the post-merger scenario. The discovery of the magnetic field of the FS~CMa star, IRAS~17449+2320, strengthened this hypothesis \citep{Korcakova2022}, as theoretical models \citep{Schneider2019} have shown that strong magnetic fields could be generated during the merger process, along with a large amount of gas and dust that is released, creating a shell around the central star. At the early stages, these resulting objects are fast rotators that slow down due to internal restructuring and magnetic braking, and become slow rotators in approximately 10$^2$ years. The resulting objects are also rejuvenated. IRAS~17449+2320 shows a peculiar large space velocity, which could also be a result of the merger process \citep{Dvorakova2024}. This star is the topic of our study.\smallskip

\cite{Korcakova2022} found that some metallic lines in the spectrum of IRAS~17449+2320 showed Zeeman splitting, such as \ion{Fe}{I}, \ion{Fe}{II}, \ion{Si}{II}, \ion{Cr}{I}, \ion{Cr}{II}, \ion{Ti}{II}, \ion{C}{I}, \ion{N}{I}, and \ion{O}{I}. They measured the magnetic field modulus of each line, and obtained an average value of the mean magnetic field modulus of $|B|$= 6.2$\pm$ 0.2~kG. They set the spectral type of this target to A0 or B9 due to the lack of iron lines that should appear in later stellar types. They estimated an upper limit of the effective temperature to be 11,040~K, the $\log g$ to 4.1, and a rotational velocity, v$\sin i$, of 9.1 km s$^{-1}$ using the Python code PYTERPOL \citep{Nemravova2016}. This fitting also provides some stellar parameters of a putative secondary with a T$_{eff}$ of around 51,000~K, log g of 4.0 and a rotational velocity of 800~km s$^{-1}$. In this specific case, this putative secondary corresponds to an additional hot continuum source.\smallskip

The occultation of the central star due to the clouds of gas and dust \citep{DeWinter1997} makes the determination of the stellar parameters of these objects complicated. This may explain the significant discrepancies among the stellar parameters. \cite{Condori2019} report the spectral type of IRAS~17449+2320 to be A0-A2. They provided three different temperature values based on three different methods (references therein), 9,200 $\pm$ 300 K, 9,500 $\pm$ 500 K, and 17,000 $\pm$ 700 K.\smallskip

Both studies also estimated the mass of this star using different evolutionary tracks. \cite{Condori2019} suggested that the star is in the main sequence or close to the end, with a M$_{ZAMS}$ = 2-3 M$_{\odot}$ (Fig. 6 and references therein), and \cite{Korcakova2022} estimated the mass of the star to be 2.4 M$_{\odot}$ (Fig. 6 and references therein).\smallskip

The most recent study performed for IRAS~17449+2320 was carried out by \cite{IRASProceedings2025}. In this paper, they estimated the effective temperature range between 9\,800 $\pm$ 300 K and a $\log g$ $\sim$ 3.8. They also determined a 36-day cycle. They obtained this cycle from the variability in the ratio of the flux minima on the red and blue sides of H$_\alpha$. They report a detection of a magnetic field based on Zeeman splitting of infrared \ion{N}{I} lines, confirming the results of \cite{Korcakova2022}, and report magnetic field measurements based on the \ion{N}{I} 8703 \AA~line, but do not provide individual numerical values or details of their method.\smallskip

As can be seen, the nature of IRAS~17449+2320, and FS~CMa stars generally, makes the study of these objects complicated. The emission contamination in their spectra does not permit a straightforward analysis of the spectral characteristics of these objects. Studying the magnetic field of IRAS~17449+2320 is of special interest as it will help us determine its stellar parameters, and therefore understand the evolutionary state of this star. Additionally, it could give insights into the origin of some other FS~CMa stars.

In this paper, we present a study of the magnetic field of IRAS~17449+2320. In Section \ref{sec:observations}, we introduce the observations used for this work. Section \ref{sec:high-res} shows the first characterization of the high-resolution polarized spectra of this object. In Section \ref{sec:Measurements}, we introduce all the measurements obtained from the spectra and used to perform the analysis of the magnetic field. The determination of its rotational period is explained in Section \ref{sec:period}. The origin of this object is discussed in more detail in Section \ref{sec:discussion}.

\begin{figure*}[ht]
    \centering
    \includegraphics[width=\textwidth]{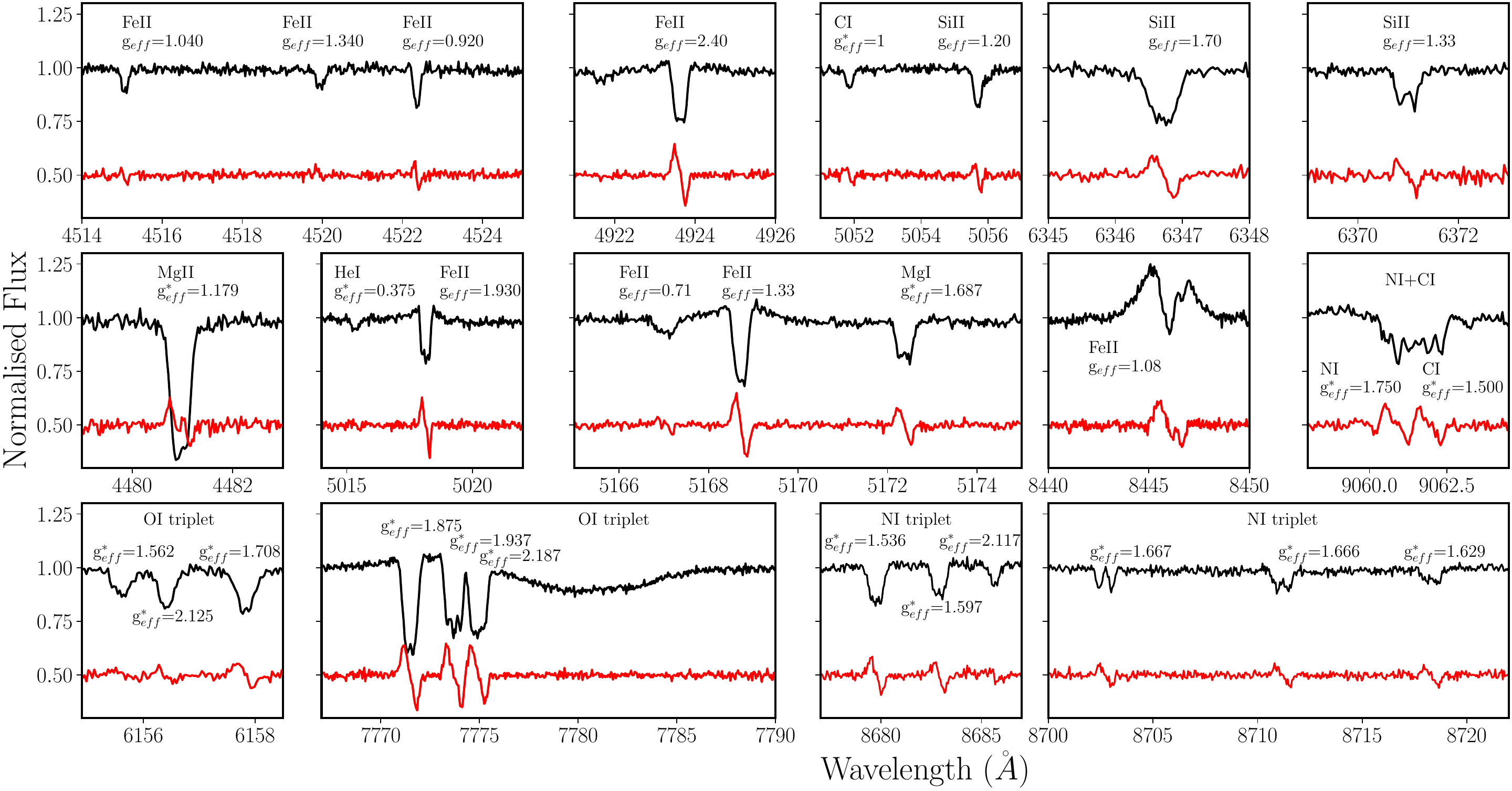}
    \caption{Spectrum (black) and circular polarization, Stokes $V$/$I_c$, (red) of multiple metallic lines throughout the wavelength range of IRAS~17449+2320. The upper panels show individual metallic lines, some of them showing Zeeman splitting. The middle panels show lines with combined emission and absorption profiles and V Stokes profiles with peculiar shapes. The lower panels show some triplets. In each panel, the Landé factors are specified; g$_{eff}$ are the measured values, and g$_{eff}^{*}$ are the theoretical values obtained from VALD3 \citep{Piskunov1995, Ryabchikova2015}.}
    \label{fig:spectrum+vstokes}
\end{figure*}

\section{Observations}\label{sec:observations}
There are 12 spectropolarimetric (Stokes V) observations for IRAS~17449+2320 taken over the last 19 years. All of them were taken using ESPaDOnS at the Canada-France-Hawaii Telescope (CFHT). These observations are or will be publicly available in the CFHT archive, with the six most recent ones obtained during semesters 24A, 24B and 25A\footnote{CFHT program IDs: 24AC019, 24BC021, 25AC12.}. ESPaDOnS is an optical echelle spetropolarimeter which consists of a bench-mounted spectrograph and a polarimetric module placed at the Cassegrain focus. It has a resolving power of $\sim$ 65,000 that covers the entire optical range (from 370 to 1,050 nm).

Each polarimetric observation consists of four subexposures with a different rotation of the instrument's Fresnel rhombs. This process produces an unpolarized intensity spectrum (Stokes I), a circular polarized (Stokes V) spectrum, and two diagnostic null (N) spectra. These null spectra are used to identify artificial signals in the Stokes V spectra. We use the polarized Stokes V spectra to detect and characterize the stellar photospheric magnetic field, as these spectra are sensitive to the Zeeman effect \citep[e.g][]{Donati2009}, as can be seen in Fig. \ref{fig:spectrum+vstokes}, where an example of selected regions of a typical Stokes V spectrum are shown. The log of the observations is shown in Table \ref{tab:b}. In addition to the nine Stokes V observations for this star, two additional Stokes Q and two Stokes U observations were taken in 2006 and 2008 (Appendix \ref{app:1}).\smallskip

\section{First characterization of the high-resolution polarized spectrum of IRAS~17449+2320}\label{sec:high-res}

\cite{Korcakova2022} worked with multiple spectra with high and low resolution (Table 2 therein) and different wavelength ranges to obtain an overview of the characteristics of IRAS~17449+2320.\smallskip

One of the most characteristic features of this object are the Balmer lines and the emission pollution they present in the core of the line. They observed 17 Balmer lines, H$_{\theta}$ being the first component showing weak emission. This emission gets stronger and more complex in the lower members of the Balmer series. H$_{\alpha}$ is of special interest as its emission is five to ten times stronger than the continuum, but still shows well defined absorption wings. The Paschen series are also seen from Pa$_{\epsilon}$, and they present emission in their cores too.

There are also multiple \ion{He}{I} lines. Most of them are in absorption and present no significant variability with time. However, \ion{He}{I} $\lambda$ 5976, 6678 \AA~ present their wings in emission, and their profiles vary with time. The \ion{He}{I} lines are the ones used to constrain the spectral type of this object. The \ion{He}{I} lines are weak and point to a late B-type star. The presence of narrow metallic lines, as \ion{Fe}{II}, \ion{Ti}{II}, and \ion{Mg}{II} is consistent with this spectral type.

IRAS~17449+2320 only shows forbidden emission lines of [\ion{O}{I}] $\lambda$ 6300, 6364 \AA. The rest of the neutral oxygen lines present as triplets show Zeeman splitting and an emission component in the \ion{O}{I}~7771~\AA ~triplet. \cite{Korcakova2022} also found two resonance line doublets, \ion{Na}{I} D1, D2, and \ion{Ca}{II} H, K, both of them showing broad emission overlapped with the interstellar component. In Figure \ref{fig:spectrum+vstokes_emission}, we show illustrative ESPaDOnS observations of these lines, showing no individual circular nor linear polarization.\smallskip

As illustrated in Fig. \ref{fig:spectrum+vstokes}, essentially all lines exhibit a consistent circular polarization variation across their profiles, with morphologies and amplitudes consistent with the Zeeman effect. Some lines, such as Fe~{\sc ii}\ 5018~\AA~ (Fig.~\ref{fig:spectrum+vstokes} second row second panel), or O~{\sc i}\ 7773~\AA~ (Fig.~\ref{fig:spectrum+vstokes} third row second panel) exhibit triplet-like splitting in their Stokes $I$ profiles, and the extrema of their associated circular polarization profiles are coincident with the $\sigma$ components of the triplets, as expected from the Zeeman effect. These properties strongly confirm the interpretation by \citet{Korcakova2022} that the splitting structures observed in the absorption lines result from a strong magnetic field, rather than a spectroscopic binary companion, for example.\smallskip

We also note that the polarity of these Stokes $V$ signatures remains the same in both absorption and emission lines. As discussed by \citet[][NGC~1624-2]{Wade2012} and \citet[][HD~45166]{Shenar2023}, this suggests that the circular polarization is associated with absorption lines formed in the stellar photosphere. Absorption lines are then infilled by emission produced in the circumstellar environment. If the magnetic field in the region generating emission is much weaker than the magnetic field in the photosphere, then the emission contributes negligibly to the Stokes $V$ profiles. The Zeeman effect occuring in emission will yield profiles of opposite polarity to profiles formed in absorption, for the same magnetic field orientation. Thus the similarity of the  Stokes V signatures in lines with and without emission suggests that in all cases the signatures are coming from photospheric absorption, and therefore, that the magnetic field in regions generating emission is much weaker than in the photosphere.\smallskip

The circular polarization changes over time. The Stokes $V$ profiles not only change shape and intensity, but also polarity over the course of the time series (Fig. \ref{fig:all-LSD}).\smallskip

\section{Measurements}\label{sec:Measurements}
\subsection{Least squares deconvolution}\label{sec:data_reduction}
\begin{figure*}[ht]
    \centering
    \includegraphics[width=\textwidth]{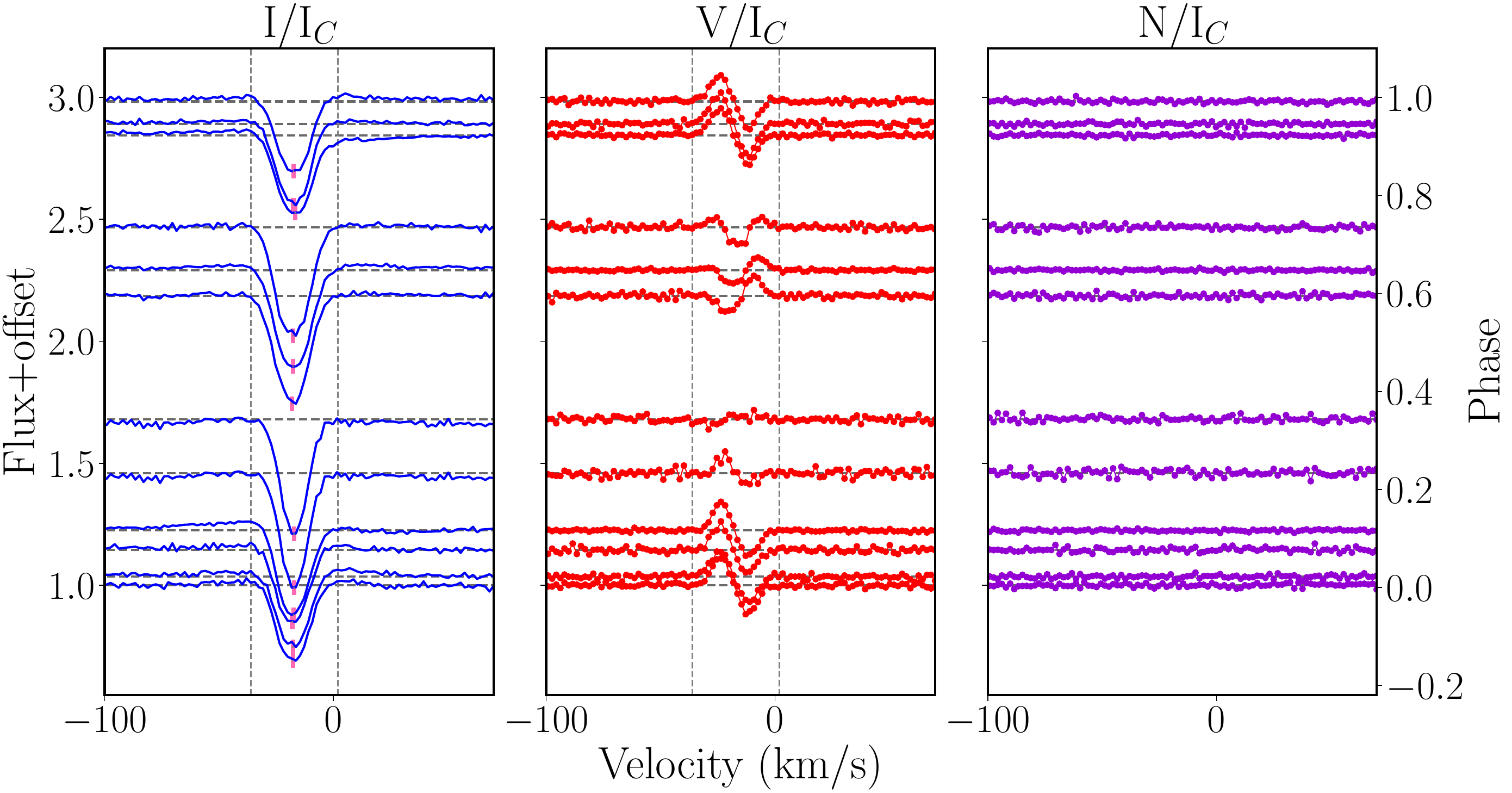}
    \caption{LSD Stokes $I$, LSD Stokes $V$, and LSD $N$ profiles, respectively, all normalized by the Stokes $I$ continuum ($I_c$). The observations are ordered and shifted vertically by phase using Eq. \ref{eq:phase}. The vertical dashed lines are the limits used to calculate B$_z$, and the pink lines mark the radial velocity of each observation.}
    \label{fig:all-LSD}
\end{figure*}

\begin{figure}[ht]
    \centering
    \includegraphics[width=\hsize]{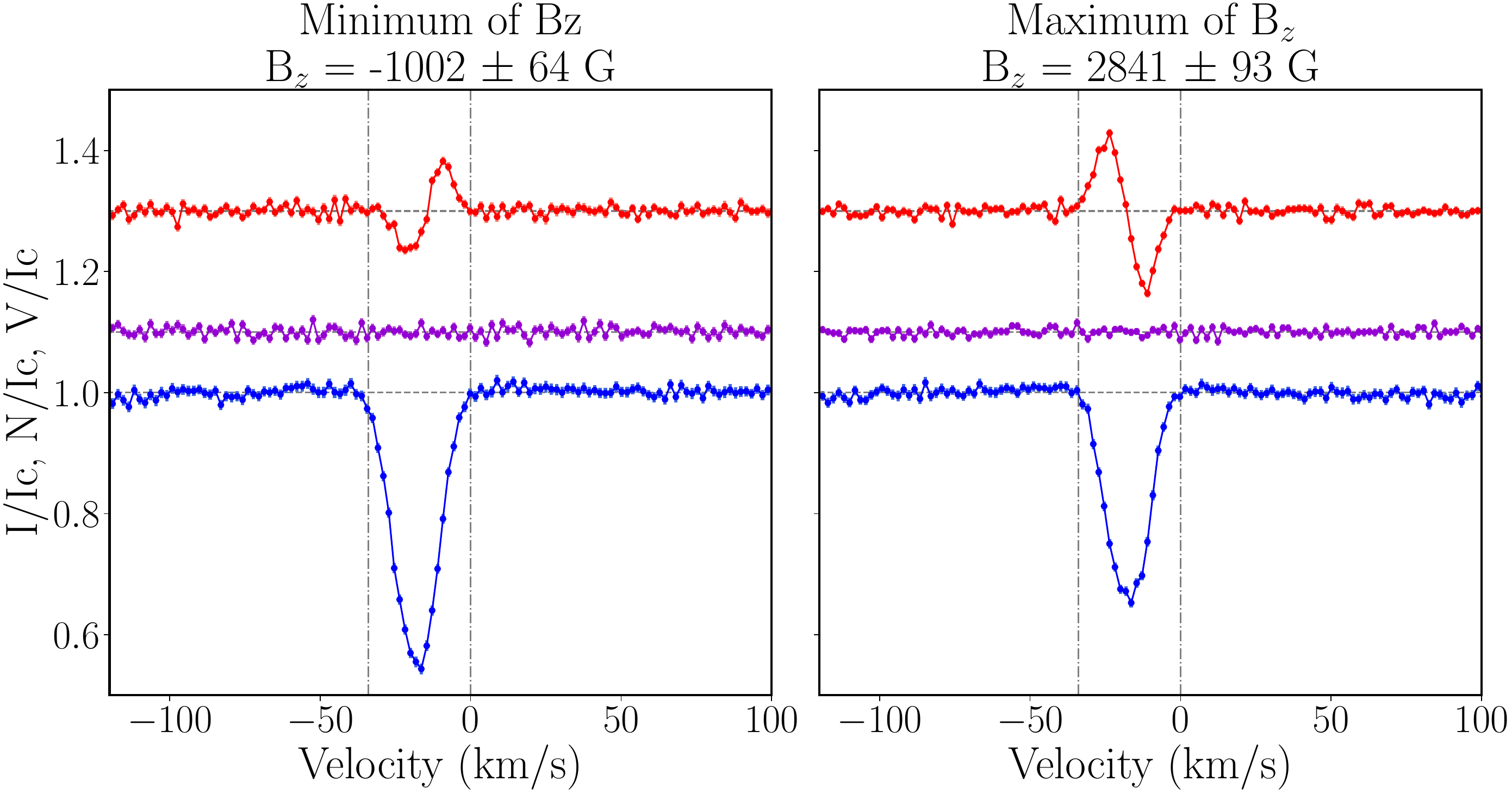}
    \caption{LSD Stokes $V$ (red), LSD Stokes $I$ (blue), and LSD $N$ (purple) profiles of two different observations with opposed phases. These three profiles are normalized by $I_c$. The figure on the left corresponds to the minimum value of B$_{z}$ calculated, and the figure on the right corresponds to the maximum. The vertical lines are the limits used to calculate B$_{z}$.}
    \label{fig:lsd-prof}
\end{figure}

We used a least squares deconvolution (LSD) \citep{Donati1997} multiline analysis procedure on each spectrum to enhance the signal-to-noise ratio (S/N) of Zeeman signatures using the SpecpolFlow routine \footnote{\href{https://github.com/folsomcp/specpolFlow}{https://github.com/folsomcp/specpolFlow}} \citep{Folsom2025}. This routine retrieves a high signal-to-noise ratio mean circular polarization profile (LSD Stokes $V$), a mean unpolarized profile (LSD Stokes $I$) and a mean diagnostic null proﬁle (LSD $N$). The line mask used for this process was obtained from the Vienna Atomic Line Database (VALD3; \citep{Piskunov1995, Ryabchikova2015}). To generate the mask, we assumed solar metallicity, an effective temperature of 11,000 K, and a surface gravity ($\log g$) of 4.0. This generated mask was iteratively cleaned and tweaked\footnote{The final tweaked mask is available on the \href{https://zenodo.org/records/19915806?token=eyJhbGciOiJIUzUxMiJ9.eyJpZCI6IjAyY2Q1Yjk5LWRlMjgtNGMwNy05YTQwLTM0YmFjZGNlYWVkMyIsImRhdGEiOnt9LCJyYW5kb20iOiI1MGFkYWRlYTZiOTZhZTJkODQwODk3MjkzYjRjMjBlZCJ9.5dAQzr08fUG5BCquqyfHE5SPXzV6IGM23uF6x5cJejkAyo7rLwAzWXo4LsiJAcQh926m7tElfPUTsdYzUHvlsg}{Zenodo repository}} \citep{Grunhut2017}, and used for every observation. Due to the peculiar nature of IRAS~17449+2320, there were only a few useful lines (157 lines out of the 510 lines of the mask) left for the LSD process. Hydrogen lines were discarded as their cores were in emission, and helium lines also presented contamination signals from the circumstellar matter. Some of the remaining metallic lines showed a hybrid emission and absorption profiles, those lines had to be discarded, even if they exhibited detectable circular polarization (second row of Figure \ref{fig:spectrum+vstokes}. Fig. \ref{fig:all-LSD} presents the full set of LSD profiles ordered by rotation phase (see Section \ref{sec:period}), Fig. \ref{fig:lsd-prof} shows a summary of the extracted LSD profiles at the maximum and minimum value of the longitudinal magnetic field.\smallskip

To obtain the LSD profiles, we set the pixel size to 1.8 km s$^{-1}$, the Landé factor to 1.14, $\lambda_0$ to 600 nm, and the depth to 0.4. These three values are used for the scaling of the LSD profiles that are plotted in Fig. \ref{fig:all-LSD}. These LSD profiles are normalized to the local continuum ($I_c$).

\begin{table*}[ht]
 \caption{\label{tab:b}Compilation of the spectropolarimetric observations and magnetic results.}
\begin{tabular}{ccccrrccccc}
 \hline\hline 
    
    \begin{tabular}[c]{@{}c@{}}Date\\ ~\end{tabular} & \begin{tabular}[c]{@{}c@{}}HJD\\ (UTC)\end{tabular} & \begin{tabular}[c]{@{}c@{}}Phase\\ ~\end{tabular} & \begin{tabular}[c]{@{}c@{}}Int. time\\ (s)\end{tabular} & \begin{tabular}[c]{@{}c@{}}B$_z$ $\pm$ $\sigma_{B_z}$\\ (G)\end{tabular} & \begin{tabular}[c]{@{}c@{}}RV $\pm$ $\sigma_{RV}$\\ (km/s)\end{tabular} & \begin{tabular}[c]{@{}c@{}}$|B|$ (G) \\ SiII 6371\AA\end{tabular} & \begin{tabular}[c]{@{}c@{}}$|B|$ (G) \\ CaI 8497\AA\end{tabular} & \begin{tabular}[c]{@{}c@{}}$|B|$ (G) \\ NI 8703\AA\end{tabular} & \begin{tabular}[c]{@{}c@{}}$\sigma_{|B|}$\\ (G)\end{tabular} \\
 \hline

    09-06-2006 & 2453895.869 & 0.586 & 720 & -1002 $\pm$ 64 & -18.06  $\pm$  0.08 & 5535 & 5800 & 5336 & 50 \\
            
    25-07-2008 & 2454673.031 & 0.105 & 2100 & 2267 $\pm$ 65 & -17.40 $\pm$ 0.12 & 4925 & 5281 & 4838 & 50 \\
            
    09-02-2012 & 2455967.132 & 0.938 & 600 & 2841 $\pm $93 & -17.51 $\pm$ 0.15 & 5395 & 5895 & 5060 & 50 \\
            
    13-02-2012 & 2455971.101 & 0.048 & 600 & 2688 $\pm$ 103 & -17.69 $\pm$ 0.19 & 5019 & 5977 & 5045 & 50 \\
            
    13-08-2017 & 2457978.740 & 0.638 & 1600 & -598 $\pm$ 38 & -17.59 $\pm$ 0.07 & 5408 & 5849 & 5451 & 50 \\
            
    26-05-2021 & 2459361.086 & 0.915 & 1800 & 2319 $\pm$ 73 & -16.72 $\pm$ 0.15 & 5155 & 5884 & 4981 & 50 \\
            
    01-01-2024 & 2460311.175 & 0.249 & 700 & 788 $\pm$ 101 & - & - & - & - & - \\
            
    05-01-2024 & 2460315.157 & 0.332 & 700 & -449 $\pm$ 105 & -17.21 $\pm$ 0.09 & 4373 & 4388 & 4615 & 50 \\
            
    23-08-2024 & 2460546.058 & 0.726 & 728 & 87 $\pm$ 69 & -17.72 $\pm$ 0.12 & 5607 & 5959 & 5410 & 50 \\ 
            
    06-04-2025 & 2460772.065 & 0.984 & 728 & 2635 $\pm $98 & -17.44 $\pm$ 0.18 & 5072 & 5578 & 5567 & 50 \\
            
    07-04-2025 & 2460773.011 & 0.010 & 728 & 2614 $\pm$ 94 & -17.68 $\pm$ 0.18 & 5062 & 5338 & 5644 & 50 \\
            
    08-04-2025 & 2460774.994 & 0.065 & 728 & 2247 $\pm$ 104 & -17.97 $\pm$ 0.20 & 5004 & 5155 & 5878 & 50 \\
 \hline
\end{tabular}
 \tablefoot{Observations were taken with ESPaDOnS at CFHT. Each column shows the civil date of the observation, the Heliocentric Julian Date, the phase of the observation as calculated using Eq. \ref{eq:phase}, the integration time, the longitudinal magnetic field with its error, the radial velocity obtained from the LSD $I$ profiles and its error, the magnetic field modulus obtained from the Zeeman splitting of the specific \ion{Si}{II}, \ion{Ca}{I}, and \ion{N}{I} lines, and the error for the magnetic field modulus (see Section \ref{sec:b} for an explanation on the estimation of this error).}
\end{table*}

\subsection{Longitudinal magnetic field}\label{sec:bz}
The detection of the Stokes $V$ signal is the key indicator of a photospheric magnetic field in a star. In order to calculate the longitudinal magnetic field, we are using the first moment method \citep{Donati1997, Wade2000b} on the LSD Stokes $V$ profile of each observation. It can be obtained by evaluating:

\begin{equation}
    B_{z} = \frac{1}{g_\mathrm{eff} \lambda_{\rm B, cte} \lambda_o c}\frac{\int vV(v)dv}{\int (1-I(v))dv}
\label{eq:bz}
\end{equation}

\noindent where $I$ and $V$ are normalized by the continuum of $I$ ($V/I_c$ and $I/I_c$), $\lambda_0$ is the wavelength of the transition in angstroms (\AA), and $\lambda_{\rm B, cte}~=~4.67\times 10^{-13} \AA^{-1} \mathrm{G}^{-1}$ is a constant for the Zeeman splitting. The speed of light, $c$, has the same units as the velocity, $v$. For LSD profiles, $g_{\rm eff}$ is the normalizing Land\'e factor and $\lambda_0$ is the normalizing wavelength (1.14 and 6\,000 \AA\ in our study). The extent of the LSD Stokes $V$ parameter is delimited by the LSD Stokes $I$ profile. The measure of the mean longitudinal magnetic field was calculated using the Python software SpecpolFlow. The integration of the B$_z$ equation was carried out within $\pm$18~km~s$^{-1}$ of the center of gravity of the Stokes I profile. The center of gravity of Stokes I, used as $v = 0$ in Eq. \ref{eq:bz}, was calculated from the velocity range -38 to 3~km~s$^{-1}$. These values are the same in each observation.

It is also necessary to calculate the False Alarm Probability (FAP) of the Stokes V signal \citep{FAP1992}. For a definitive detection, FAP < 10$^{-5}$, and 10$^{-3}$ for a marginal detection. From our dataset, we only have one marginal observation, the night of the 5 January 2024. The rest are definite detections. The FAP and the longitudinal magnetic field calculation are done over the same velocity range. All the values of the longitudinal magnetic field and its uncertainty (calculated with an error propagation of Eq. \ref{eq:bz}) are presented in Table \ref{tab:b}. The longitudinal magnetic field shows a stable variation over time with a maximum value of 2841 $\pm$ 93 G and a minimum of -1002 $\pm$ 64~G.\smallskip

For the observation on 1 January 2024, the calculation of B$_{z}$ was slightly different. This observation finished 2 min after $\ang{;;8}$ twilight, and the whole spectrum was polluted with solar lines, which added an extra absorption line in the LSD $I$ profile. However, the LSD $V$ Stokes profile was not affected by the solar contribution. Therefore, we used the LSD $I$ profile from the observation of 5 January 2025, in combination with the Stokes V profile from 1 January 2024, to infer B$_{z}$. The error introduced in this process is not significant, as the line depth and radial velocity do not vary much from night to night (Table \ref{tab:b}). Due to the solar pollution, the 5 January 2024 observation was only used for the B$_{z}$ analysis, but not for the $|B|$ analysis.

\subsection{Period analysis}\label{sec:period}
The estimation of the rotational period of IRAS~17449+2320 is key for the determination of other stellar parameters for this object, especially after considering the uncertainty around these values due to the peculiar nature of FS~CMa stars (see Section \ref{sec:Introduction}). We performed a time series analysis to determine the periodicity of the B$_{z}$ dataset, which would be the rotational period in the context of the Oblique Rotator Model (ORM) \citealt{Monaghan1973}; \citealt{Mestel2005}.\smallskip

\begin{figure}
    \centering
    \includegraphics[width=\hsize]{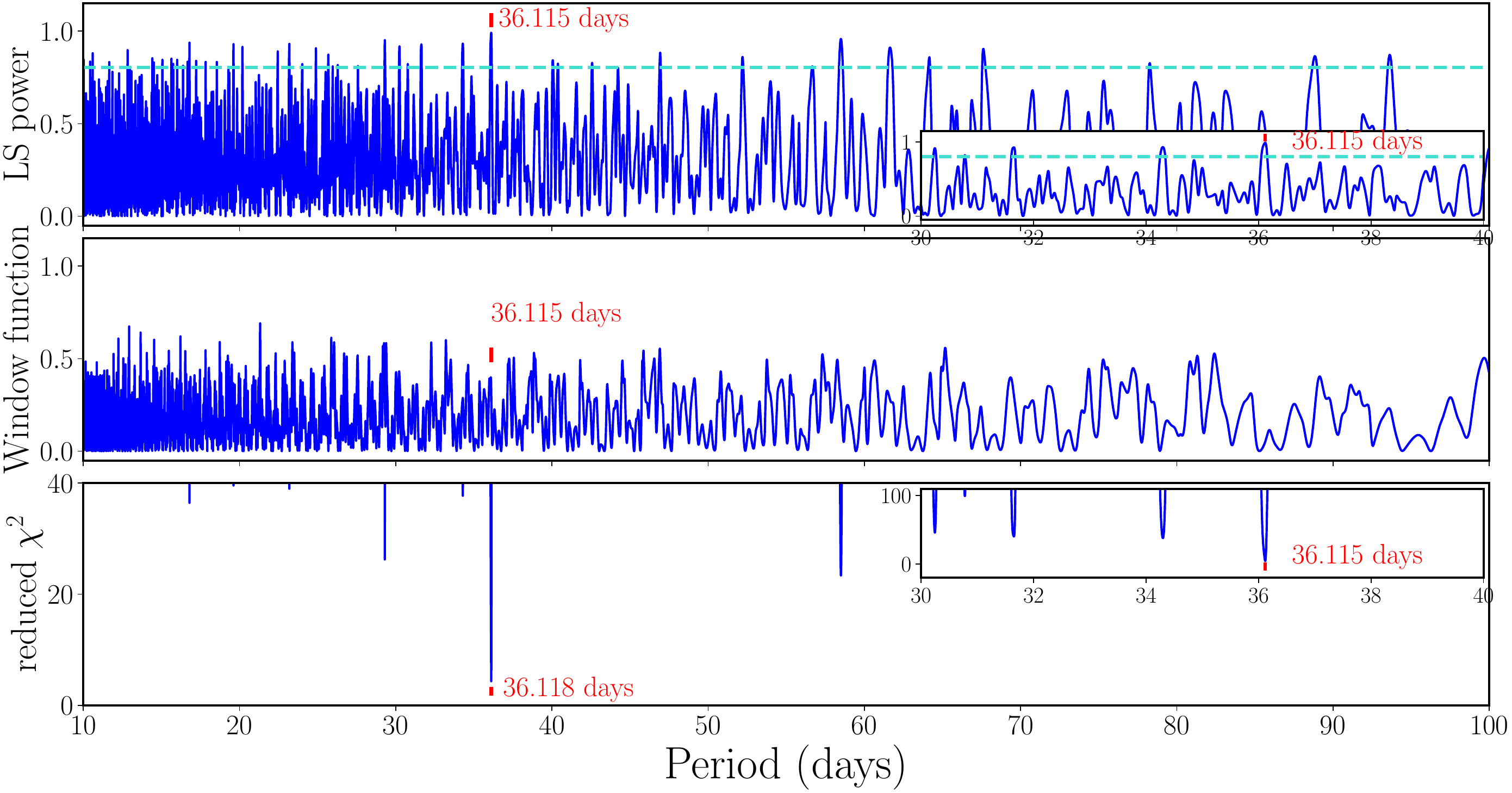}
    \caption{Periodogram of B$_{z}$ dataset from 10 to 100 days to show the most significant period values. The subplot presents a zoom in on the periodogram from 30 to 40 periods to show the highest probability period: 36.11 days. The middle panel shows the estimate of the LS periodogram window function inferred from the null spectrum LSD longitudinal field measurements.}
    \label{fig:periodogram}
\end{figure}

At the beginning, we estimated a period of 13 years with the first six spectropolarimetric observations available in the archive. Subsequent observations taken during semesters 24A, 24B and 25A showed that the period was in fact considerably shorter, with variability on timescales around one month. Therefore, we performed a Lomb-Scargle (LS) periodogram \citep{Lomb1976}, using the Python tool astropy \citep{timeseries2015}, over the dataset between 1 day to 100 days in order to have a wide range of possible periods, as the most likely periods, based on the periodogram results, are in between these values (Fig. \ref{fig:periodogram}, upper panel). We extended the range of the period search down to 1 day and up to 1\,000 days, but we did not find any more significant peaks in the periodogram. An estimate of the window function is shown in Fig. \ref{fig:periodogram}, computed from a LS analysis of the LSD $N$ B$_z$ values (measured in the same was as the Stokes $V$ B$_z$ values). We note specifically that the 36.11-day peak is absent in the window function.\smallskip 

The period with the highest probability is 36.115 days (marked with a red line). However, there are other values retrieved by the LS analysis that also present a high probability, i.e., 61.64 days or 30.78 days. We performed a reduced $\chi^{2}$ calculation by fitting a sinusoidal function with a fixed period, but free phase, amplitude, and zero-point to obtain which rotational value provides the best fit to the data points. The period that minimizes this reduced $\chi^{2}$ to a value of 4.25 is 36.118 days, as seen in the bottom panel of Fig. \ref{fig:periodogram}. The next best period is 58.60 days, with a reduced $\chi^{2}$ that is five times larger at 23.27. There is a small difference between the period provided by the LS periodogram and the reduced $\chi^{2}$ calculation, 36.11 and 36.12 days. However, this difference is compatible with the error considered by the LS analysis of 0.01 days, which makes both values to be in agreement. Moreover, the 36.11-day period is the only value that provides a proper fit for the longitudinal magnetic field, the magnetic field modulus (Fig. \ref{fig:mag-mod}), and the equivalent width variation (Fig. \ref{fig:period-test}) at the same time.

\begin{figure}[h]
    \centering
    \includegraphics[width=\hsize]{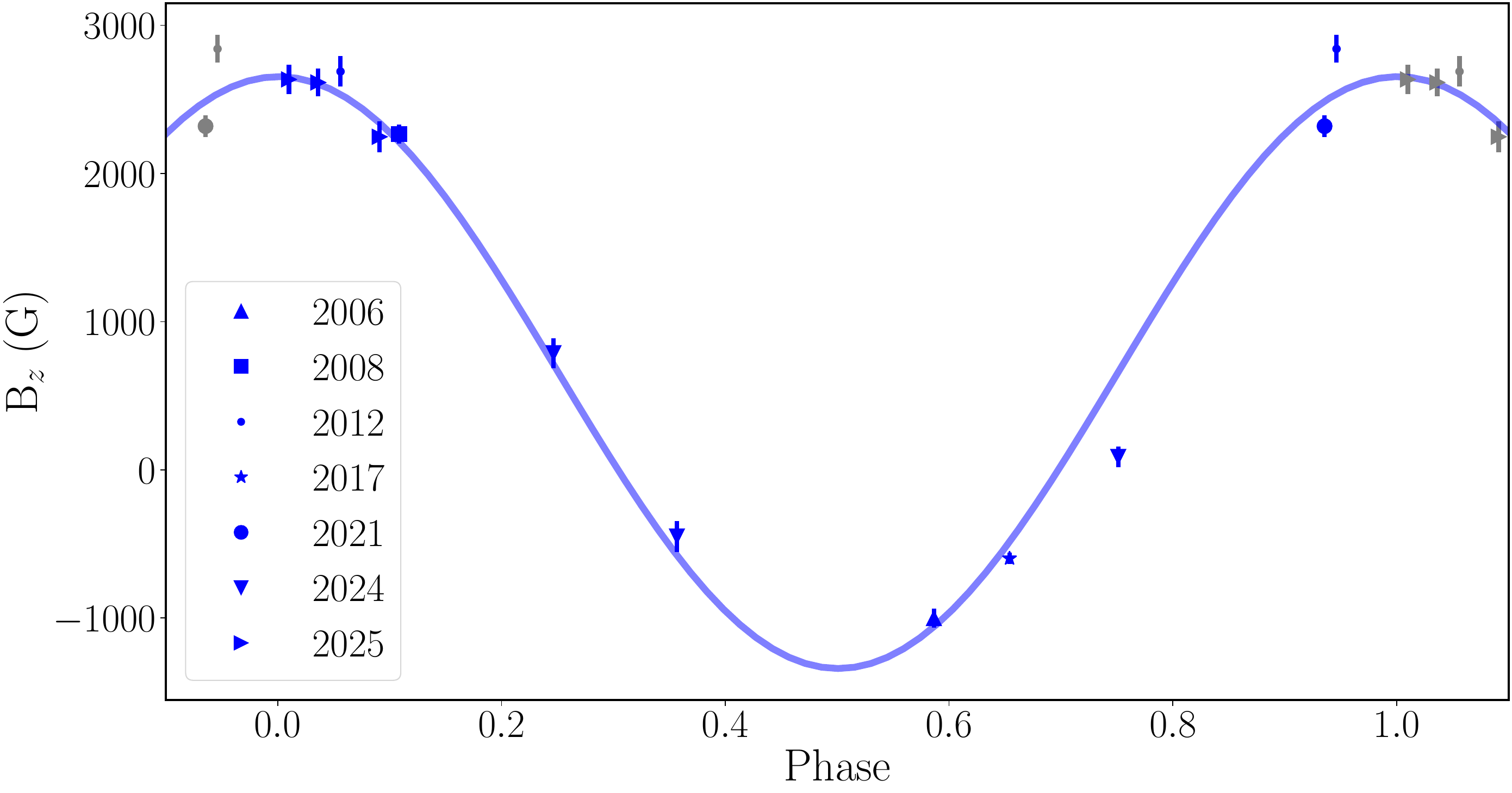}
    \caption{Longitudinal magnetic field phased using a period of 36.11 $\pm$ 0.01 days. Observations from different years are plotted with different symbols.}
    \label{fig:bz-ls}
\end{figure}

In Figure \ref{fig:bz-ls}, the sinusoidal fit corresponding to the most likely period of 36.11 days, as determined by the LS calculation, is plotted over the longitudinal magnetic field dataset. The phases are calculated with Eq. \ref{eq:phase}. There is some dispersion in the maximum values of the longitudinal magnetic field, which introduces some uncertainty in the zero-point of the ephemeris, HJD$_0$. This dispersion is likely related to the variable emission of the star. The LSD $I$ profiles of these maximum values show some weak emission features in their profiles, which explains the dispersion of the data. We chose the maximum of the B$_z$ sinusoidal fit as the Zero-point ephemeris, HJD$_0$=2455971.173. The magnetic ephemeris adopted is therefore

\begin{equation}
    HJD = 2455971.173 (0.0058) + E\times36.11 (0.01)
\label{eq:phase}
\end{equation}

\noindent where $E$ is the epoch and $HJD$ is the Heliocentric Julian Date.\smallskip

However, this analysis was performed with little data (12 observations in total). Therefore, it was important to check if this rotational period was consistent with other datasets, specifically the magnetic field modulus (see Sect. \ref{sec:b}) and the equivalent width (see Sect. \ref{sec:ew}) variation. We checked the period of 36.11 days by phasing these other datasets with this period, and clarifying they varied coherently. It can be seen in Fig. \ref{fig:mag-mod} and Fig. \ref{fig:period-test} that all the sets are in relatively good agreement with the 36.11 days period. A similar period for this object was reported by \cite{IRASProceedings2025}.

\subsection{Magnetic field modulus}\label{sec:b}

\begin{figure}[ht]
    \centering
    \includegraphics[width=\hsize]{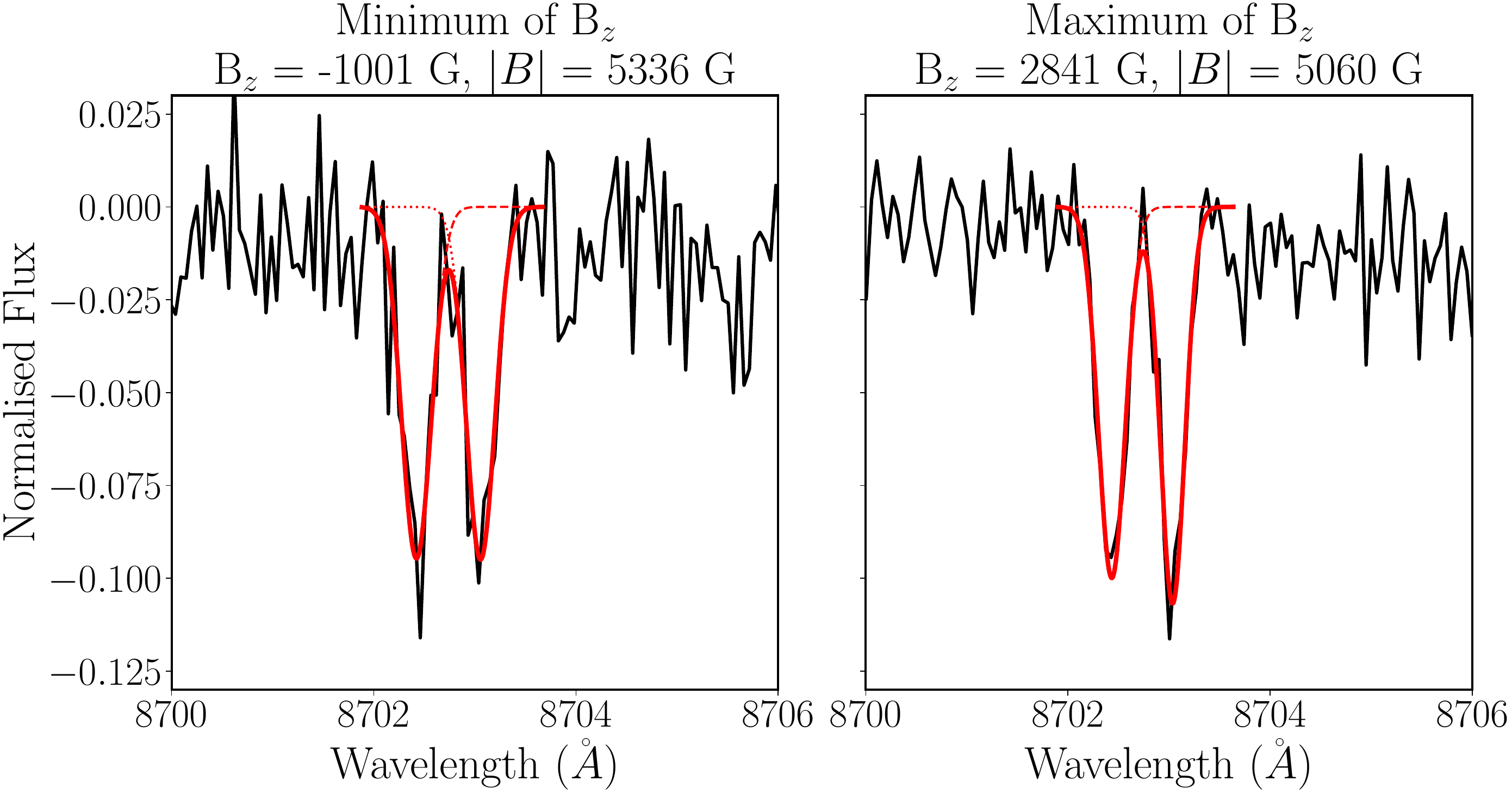}
    \includegraphics[width=\hsize]{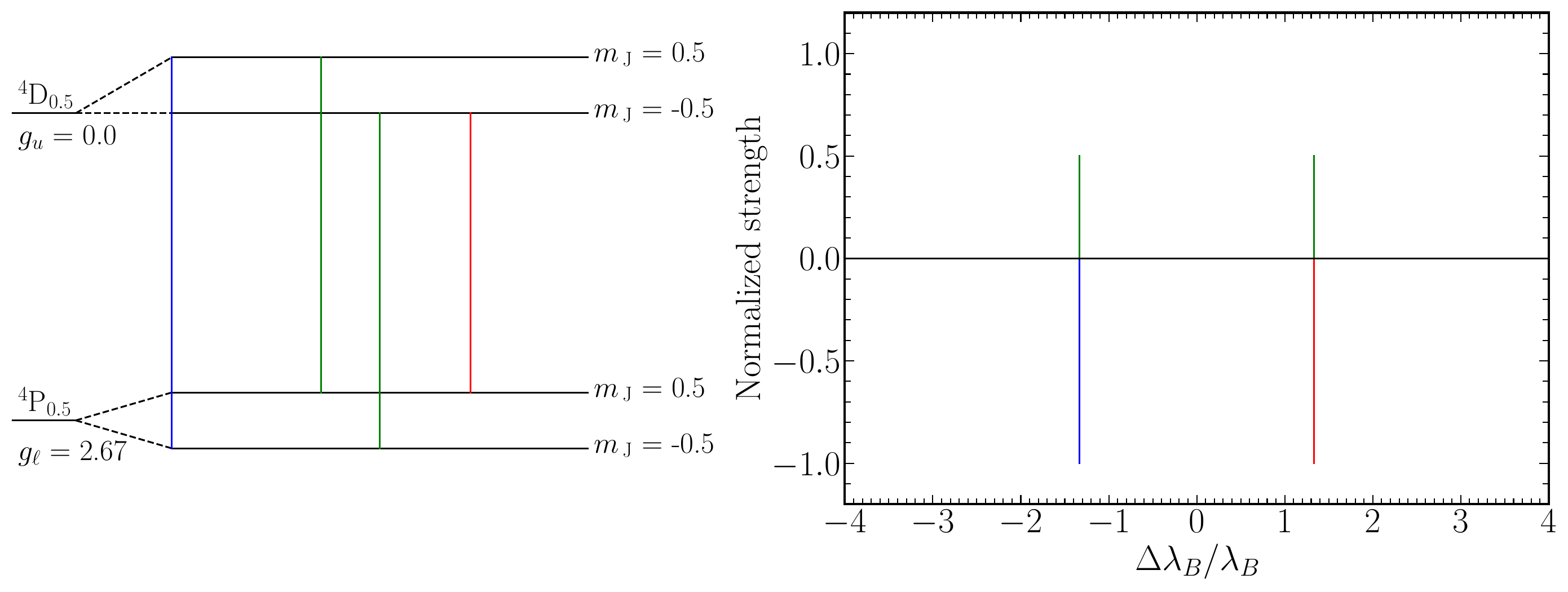}
    \caption{Gaussian fit to the two Zeeman split components of NI 8703\AA. The dotted and dashed lines show the fitting for the individual components of the line, and the solid line shows the combined two-component model of the line. The panel on the left shows the splitting for the minimum value of B$_{z}$, and the panel on the right shows the splitting for the maximum value of B$_{z}$. The two lower panels show the Zeeman pattern for this \ion{N}{I} line, corresponding to the transition of $^{4}P_{1/2}$ -- $^{4}D_{1/2}$.}
    \label{fig:zeeman}
\end{figure}

The magnetic field modulus was measured using three different lines that showed particularly clear Zeeman splitting: \ion{Si}{II} 6371\AA, \ion{Ca}{II} 8497\AA, and \ion{N}{I} 8703\AA. These three lines have relatively high effective Landé factors (1.3, and 1.67, for \ion{Si}{II} and \ion{Ni}{I}, respectively), or large wavelengths (\ion{Ca}{II}), and therefore, the splitting of their components is better defined than in other lines of the spectrum (Fig. \ref{fig:spectrum+vstokes}). Each component of the Zeeman split line was fitted with a Gaussian (Fig. \ref{fig:zeeman}, \ref{fig:zeeman_si}, \ref{fig:zeeman_ca}) to obtain an estimate of the central wavelength of the lines and calculate the Zeeman shift between two sigma components ($\Delta\lambda$ from Eq. \ref{eq:mag_mod}; \citealt{Adelman1974}):

\begin{equation}
    |B| = \frac{\Delta\lambda}{4.667\times10^{-13}\times g_{eff}\times\lambda_{0}^{2}}
\label{eq:mag_mod}
\end{equation}
where $\Delta\lambda$ is the Zeeman shift, i.e., half (for doublets) of the wavelength separation between the two components of the line, $g_{eff}$ is the effective Landé factor, and $\lambda_{0}$ is the theoretical wavelength of the line. The units for $\Delta\lambda$ and $\lambda_{0}$ are given in angstroms, and the units of the constant are G$^{-1}$\AA$^{-1}$.

The Zeeman splitting of the \ion{N}{I} 8703~\AA~line corresponds to the transition $^{4}P_{1/2}$ -- $^{4}D_{1/2}$. The \ion{Si}{II} 6371~\AA~line shows the transition $^{2}S_{1/2}$ -- $^{2}P_{1/2}$, and \ion{Ca}{II} 8497~\AA~shows the transition $^{2}D_{3/2}$ -- $^{2}P_{3/2}$. The Zeeman patterns for \ion{Si}{II} and \ion{Ca}{II} are shown in Appendix \ref{app:quantum}; Fig. \ref{fig:zeeman_si}, \ref{fig:zeeman_ca}. The \ion{Ca}{II} line shows a more complex transition pattern than the other two lines we used to measure the magnetic field modulus. 

\begin{figure}[ht]
    \centering
    \includegraphics[width=\hsize]{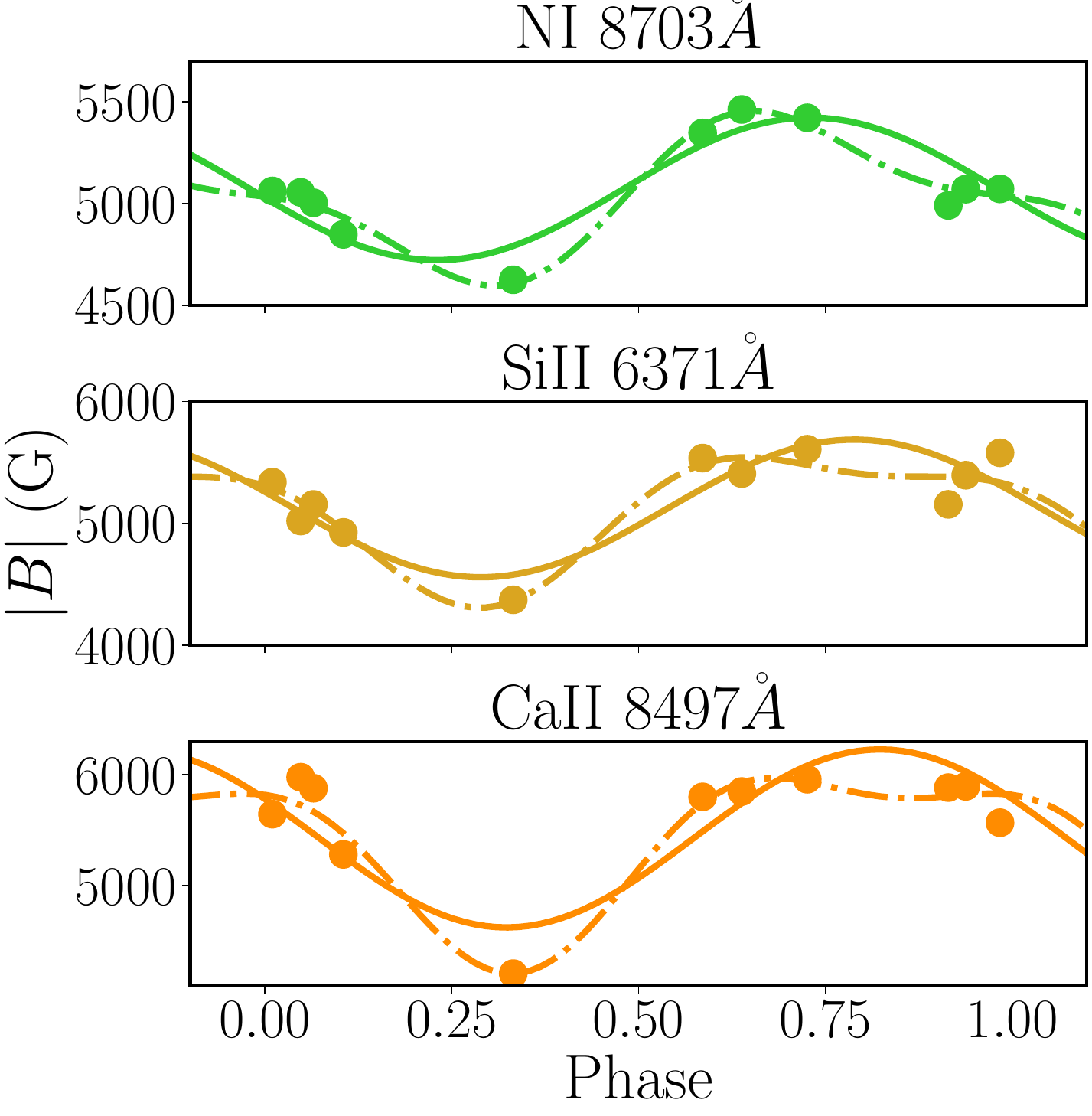}
    \caption{Magnetic field modulus obtained from three different lines showing Zeeman splitting. The data are phased with a rotational period of 36.11 days. The solid line corresponds to a first order sinusoid function, and the dashed line to a second order sinusoid function.}
    \label{fig:mag-mod}
\end{figure}

The values of the magnetic field modulus range from $\sim$6\,000~G from the \ion{Ca}{II} line to $\sim$5\,000~G for \ion{N}{I} (Table \ref{tab:b}). The discrepancy in the magnetic field modulii inferred from these three different lines is a natural consequence of the very approximate nature of the field modulus measurement approach and is expected to be resolved through detailed modeling of the spectrum using magnetic spectrum synthesis (in progress). The magnetic field modulus is shown in Fig. \ref{fig:mag-mod}, for these three lines, phased with a period of 36.11 days. We initially fitted these three datasets with a first-order sinusoidal function to verify the rotational period derived from the longitudinal magnetic field and to infer some information about the magnetic configuration. Because the first-order fit was less effective than that of the longitudinal magnetic field (Fig. \ref{fig:mag-mod}), we applied a second-order sinusoidal function. The second-order function seems to be a better fit, which is expected for a dipolar magnetic field configuration. Still, the number of data points is insufficient to conclude whether a second-order function is sufficient to describe the underlying stellar variability.\smallskip 

The estimation of the error is based on the scatter of the observations around the mean variation defined by the best-fit second-order curve discussed above. The error adopted for the magnetic field modulus for the three different lines is 50~G for each observation, noting that systematic errors in the fitting of the Zeeman components yield a larger scatter than is explained by simple noise propagation, as explained by \cite{Mathys1997}. These systematic errors are approximately the same for each line and are rather weakly dependent on the signal-to-noise ratio of the observations. These uncertainties are approximate and will be revisited as additional observations are obtained and the phase variation is better defined. \smallskip

The observation at a $\phi$=0.3 (the observation on the 05-01-2025) shows a significant decrease compared to other values. This spectrum has a much lower S/N than the rest. Therefore, the Zeeman splitting is not as well defined as in the other observations, and obtaining a proper Gaussian fit is harder.\smallskip

Additionally, there is a relatively small field-strength contrast on the star's surface, both globally and within any visible hemisphere (Fig. \ref{fig:mag-mod}). The relative change of the magnetic field modulus is 15\% for \ion{N}{I}, 24\% for \ion{Si}{II}, and 28\% for \ion{Ca}{II}. By comparing $|B|$ and B$_{z}$, it is noticeable that the magnetic field modulus is a factor of a few times the longitudinal field, which is consistent with a globally organized magnetic field, as has been established in many A, B, and O-type stars.

\subsection{Equivalent widths}\label{sec:ew}
\begin{figure}[ht]
    \centering
    \includegraphics[width=\hsize]{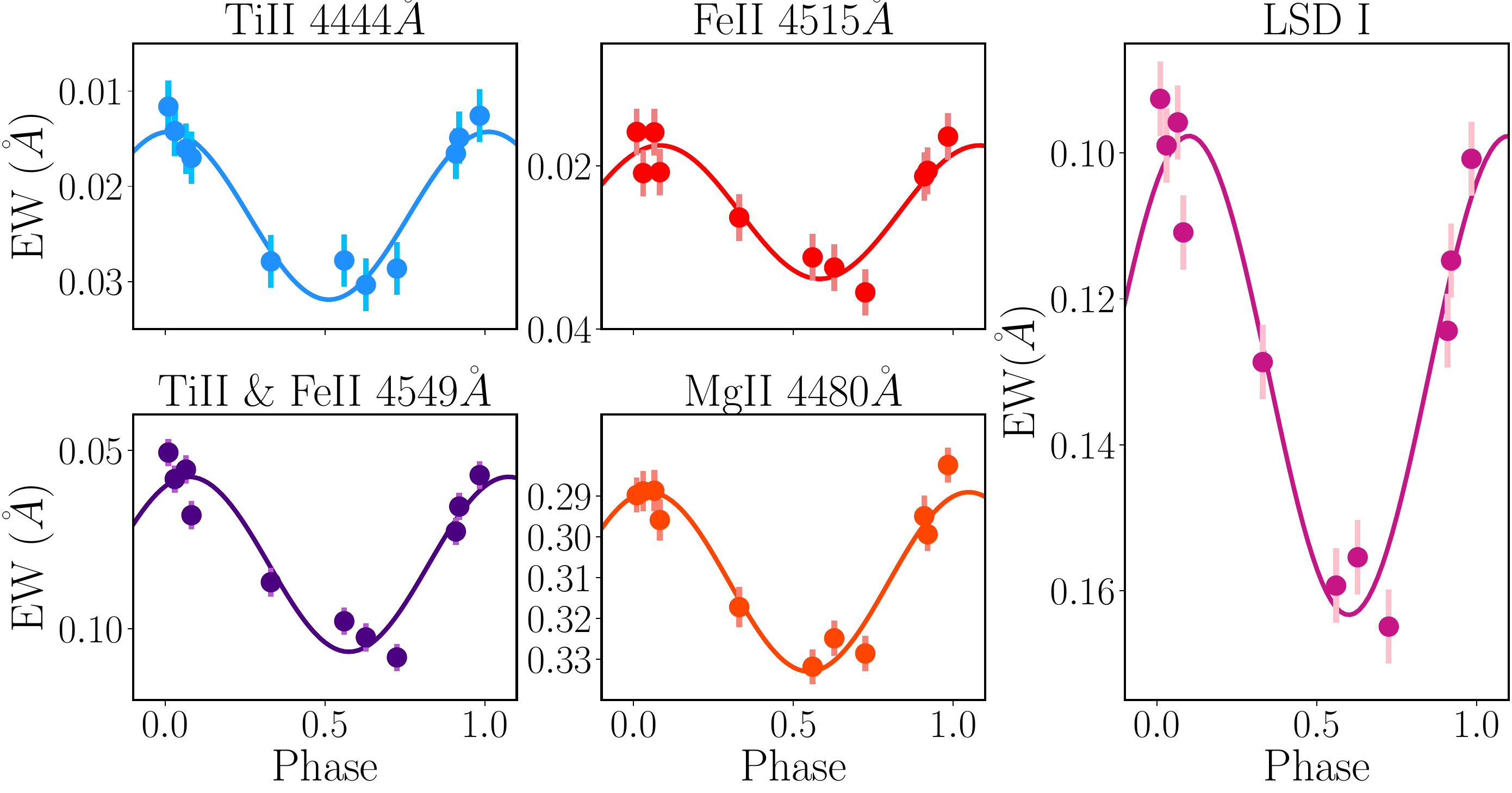}
    \caption{Plots of the equivalent width variation of some metallic lines and the LSD Stokes $I$ profile. All the datasets are phased with the 36.11 days period. The y-axis is inverted for visualization purposes.}
    \label{fig:period-test}
\end{figure}

We noticed that the equivalent width (EW) of multiple metallic lines throughout the spectrum changes with time. Some of these lines display stronger EW variation than others. Therefore, we selected a few lines that display greater variability with time: \ion{Ti}{II} 4444~\AA, \ion{Fe}{II} 4515~\AA, a blended line of \ion{Ti}{II} and \ion{Fe}{II} 4549~\AA, and \ion{Mg}{II} 4481~\AA. In addition, the Stokes $I$ LSD profiles also show an equivalent width variation. These equivalent widths were calculated using the trapezoidal rule approximation, and the results are presented in Table \ref{tab:ew} and Fig. \ref{fig:period-test}, phased with the 36.11 days period. The four metallic lines and the LSD Stokes $I$ profile show a periodic variation following a first order sinusoid which is compatible with the rotational period of the star. All of them present their maximum EW values at the minimum of the longitudinal magnetic field (Fig. \ref{fig:lsd-prof}).\smallskip

\begin{table*}[ht!]
\caption{\label{tab:ew}Compilation of the equivalent width variation of some metallic lines.}
\begin{tabular}{ccccccc}
 \hline\hline

   \begin{tabular}[c]{@{}c@{}}Date\end{tabular} & \begin{tabular}[c]{@{}c@{}}Phase\end{tabular} & \begin{tabular}[c]{@{}c@{}}EW $\pm$ $\sigma_{EW}$ (\AA)\\ TiII 4444\AA\end{tabular}  & \begin{tabular}[c]{@{}c@{}}EW $\pm$ $\sigma_{EW}$ (\AA)\\ FeII 4515\AA\end{tabular}  & \begin{tabular}[c]{@{}c@{}}EW $\pm$ $\sigma_{EW}$ (\AA)\\ TiII+FeII 4549\AA\end{tabular} & \begin{tabular}[c]{@{}c@{}}EW $\pm$ $\sigma_{EW}$ (\AA)\\ MgII 4481\AA\end{tabular} & \begin{tabular}[c]{@{}c@{}}EW $\pm$ $\sigma_{EW}$ (\AA)\\ LSD $I$ Stokes\end{tabular} \\

    \hline
    
    09-06-2006 & 0.586 & 0.0278 $\pm$ 0.0028 & 0.0312 $\pm$ 0.0028 & 0.0979 $\pm$ 0.0038 & 0.3318 $\pm$ 0.0042 & 7.965 $\pm$ 0.255 \\
            
    25-07-2008 & 0.105 & 0.0170 $\pm$ 0.0027 & 0.0207 $\pm$ 0.0029 & 0.0682 $\pm$ 0.0040 & 0.2958 $\pm$ 0.0051 & 5.544 $\pm$ 0.255 \\
            
    09-02-2012 & 0.938 & 0.0149 $\pm$ 0.0027 & 0.0206 $\pm$ 0.0028 & 0.0658 $\pm$ 0.0038 & 0.2993 $\pm$ 0.0041 & 5.737 $\pm$ 0.255 \\
            
    13-02-2012 & 0.048 & 0.0142 $\pm$ 0.0026 & 0.0209 $\pm$ 0.0028 & 0.0581 $\pm$ 0.0038 & 0.2888 $\pm$ 0.0049 & 4.948 $\pm$ 0.255 \\
            
    13-08-2017 & 0.638 & 0.0303 $\pm$ 0.0027 & 0.0324 $\pm$ 0.0029 & 0.1024 $\pm$ 0.0040 & 0.3248 $\pm$ 0.0042 & 7.771 $\pm$ 0.255 \\
            
    26-05-2021 & 0.915 & 0.0165 $\pm$ 0.0026 & 0.0213 $\pm$ 0.0030 & 0.0727 $\pm$ 0.0038 & 0.2949 $\pm$ 0.0051 & 6.219 $\pm$ 0.255 \\
            
    05-01-2024 & 0.332 & 0.0279 $\pm$ 0.0028 & 0.0263 $\pm$ 0.0029 & 0.0870 $\pm$ 0.0040 & 0.3172 $\pm$ 0.0049 & 6.433$\pm$ 0.255 \\
            
    23-08-2024 & 0.726 & 0.0286 $\pm$ 0.0027 & 0.0355 $\pm$ 0.0029 & 0.1080 $\pm$ 0.0038 & 0.3285 $\pm$ 0.0043 & 8.246$\pm$ 0.255 \\

    06-04-2025 & 0.983 & 0.0126 $\pm$ 0.0028 & 0.0164 $\pm$	0.0029 & 0.0570 $\pm$	0.0038 & 0.2824 $\pm$ 0.0042 & 5.040 $\pm$ 0.255 \\
            
    07-04-2025 & 0.010 & 0.0116 $\pm$	0.0028 & 0.0158 $\pm$	0.0029 & 0.0506 $\pm$ 0.0038 & 0.2897 $\pm$ 0.0042 & 4.629 $\pm$ 0.255 \\
            
    08-04-2025 & 0.065 & 0.0161 $\pm$	0.0027 & 0.0159 $\pm$	0.0029 & 0.0554 $\pm$ 0.0040 & 0.2887 $\pm$ 0.0051 & 4.790 $\pm$ 0.255 \\       
           
 \hline
\end{tabular}
 \tablefoot{Each column shows the civil date of the observation, the phase considering a period of 36.11 days, the equivalent width and its error of \ion{Ti}{II}, \ion{Fe}{II}, \ion{Ti}{II}+\ion{Fe}{II}, \ion{Mg}{II}, and the LSD $I$ profile.}
\end{table*}

\subsection{Radial velocity}\label{sec:RV}
The LSD $I$ profiles in Fig. \ref{fig:all-LSD} can be used for high precision measurements of the radial velocity (vertical pink lines in the center of these profiles). The radial velocity is calculated using a Gaussian fit to the LSD Stokes $I$ profile, with measurements from -18.06 $\pm$ 0.08 to -16.72 $\pm$ 0.15 km s$^{-1}$ (Table \ref{tab:b}). The variation of the radial velocity is small, approximately 1.3 km s$^{-1}$ of difference between the highest and lowest value, and a standard deviation of 0.36 km s$^{-1}$. The LSD Stokes $I$ profile with a lowest radial velocity value (-16.75 km s$^{-1}$) present a more asymmetrical core likely due to some emission contamination in the lines. Therefore, this higher dispersion in this value may be produced during the Gaussian fitting of the LSD $I$ profile rather than an actual variation of the radial velocity of the star.\smallskip

\begin{figure}[h]
    \centering
    \includegraphics[width=\hsize]{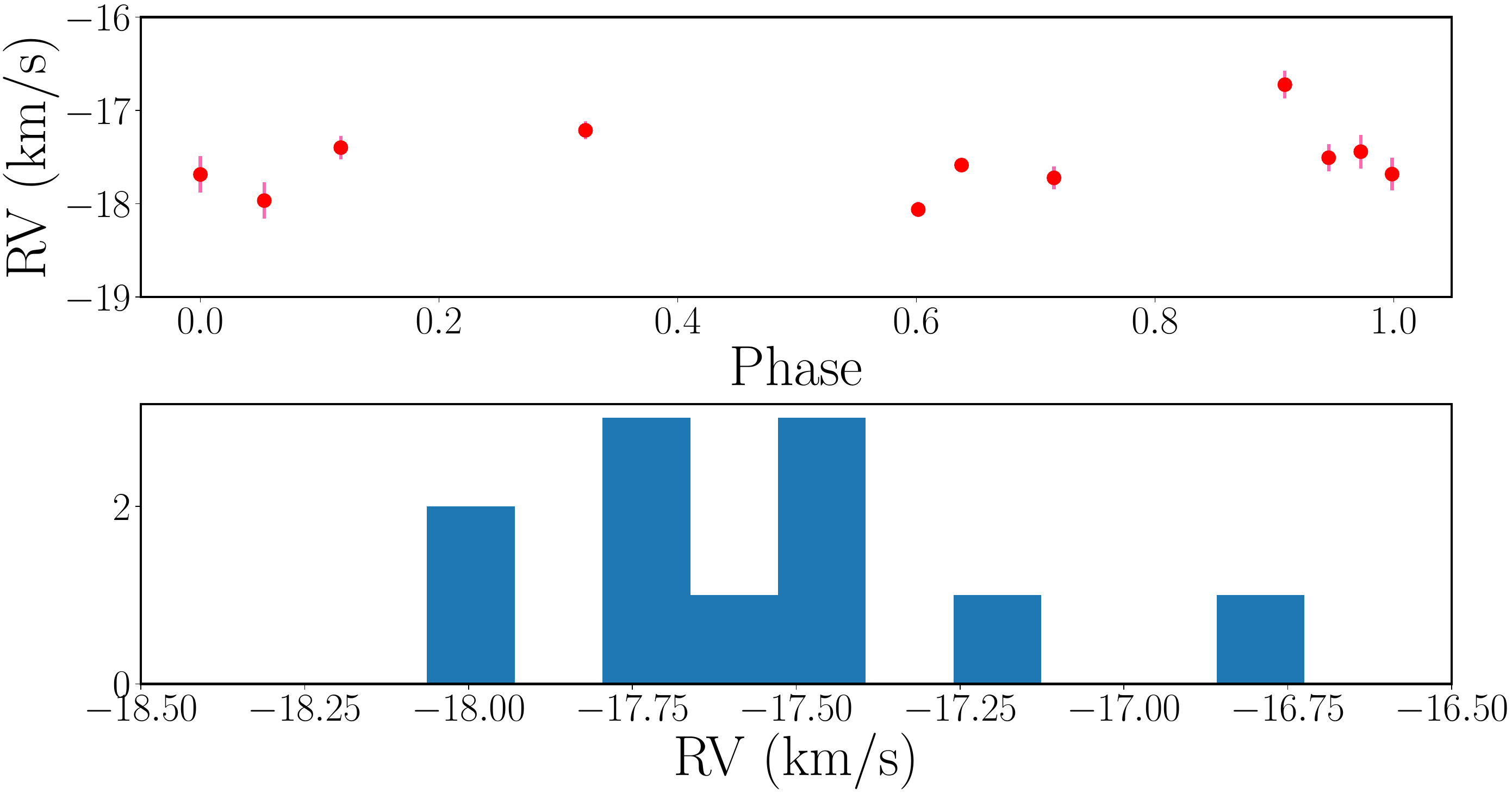}
    \caption{Radial velocity variation of the LSD $I$ profiles over time (upper panel). The bottom panel shows a histogram of these radial velocities.}
    \label{fig:rv}
\end{figure}

The radial velocity is not correlated with the magnetic field variation, and they do not vary coherently with the rotation period, as can be seen in Fig. \ref{fig:rv} (upper panel). The distribution of the measurements of the radial velocity (bottom panel of Fig \ref{fig:rv}) point to -17.6 $\pm$ 0.36 km s$^{-1}$ as the most likely value.

\section{Discussion}\label{sec:discussion}
\subsection{Characteristics of the magnetic field}

The large-scale magnetic field of IRAS~17449+2320 appears to remain stable during the last 19 years, as illustrated in Fig. \ref{fig:bz-ls}. The variation of the longitudinal magnetic field is consistent with rotational modulation. The magnetic field shows characteristics of a frozen-in organized magnetic field (Figure \ref{fig:bz-ls}), and the small variations in B$_{z, max}$ are produced by the variability of their Stokes $I$ profiles due to the circumstellar matter (as discussed in section \ref{sec:bz}).\smallskip

The sinusoidal variation of B$_z$ implies that the field has an important dipole component that is inclined to the stellar rotation axis \citep{Preston1970}. The longitudinal magnetic field shows a maximum of 2\,841 $\pm$ 93~G and a minimum of -1\,002 $\pm$ 64~G. The ratio of B$_{z,max}$ and B$_{z,min}$ is used to estimate the magnetic oblique angle, $\beta$, following \cite{Preston1970}:

\begin{equation}
    \tan \beta = \frac{1-r}{1+r} \cot i
    \label{eq:ratiobz}
\end{equation}

\noindent with $r$ being the ratio of the longitudinal magnetic field extrema \citep{Preston1967}, $i$ the inclination angle of the star, and $\beta$ the magnetic angle. The inclination angle is calculated assuming rigid rotation of the star: 

\begin{equation}
    \frac{R}{R_\odot} = \frac{T \times v \sin i}{50.6 \times \sin i}
    \label{eq:i}
\end{equation}

\noindent where T is the rotation period and the constant has units of kilometers. We consider $v \sin i$ $\sim$ 3-4 km~s$^{-1}$ (we use $v \sin i$ = 3~km~s$^{-1}$ in these calculations). This value is obtained from preliminary modeling of the spectrum using magnetic synthetic models (in progress), and it is compatible with the observed Zeeman splitting in multiple lines of this star and other metallic lines not being too wide. Other studies, such as that of \citet{Korcakova2022} or \citet{IRASProceedings2025} have reported a $v\sin i$ of 9~km~s$^{-1}$. This discrepancy between velocities could be explained by the magnetic broadening. If the magnetic broadening is not taken into consideration, the $v\sin i$ estimate is higher than the actual one. Previous studies estimated the radius of IRAS~17449+2320 to be around 2$R_\odot$ \citep{Condori2019}.\smallskip

From magnetic spectrum synthesis modeling (Bermejo Lozano  et al., in prep.), we have already estimated a range of temperature for this object, from 10\,000 K to 12\,000 K. This temperature range is consistent with other studies, for example, \cite{Korcakova2022} gave an upper limit temperature of 11\,040~K; \cite{Condori2019} provided three different temperatures through three different methods, 9\,200 $\pm$ 300, 9\,500 $\pm$ 500, and 10\,700 $\pm$ 1\,000~K; and \cite{IRASProceedings2025} reported a temperature of 9\,800 $\pm$ 300 K. We calculated the stellar radius for both 10\,000~K and 12\,000~K using the Stefan-Boltzmann law and a luminosity of $\log$ L/L$_\odot$ = 1.86 $\pm$ 0.06 from \cite{IRASProceedings2025}. The resulting radius values are presented in Table \ref{tab:i}.\smallskip

We can obtain the value of the magnetic dipole strength, using Eq. \ref{eq:mod_theoretical}, and compare it with the one we estimated from the spectra of this star. The radius, inclination angle ($i$), magnetic angle ($\beta$), and the polar strength of the magnetic dipole (B$_d$) for each temperature are shown in Table \ref{tab:i}. For this calculation, we assumed a limb-darkening coefficient ($u$) of 0.2. Due to the uncertainty of the stellar parameters, these values can be used as upper and lower limits of the magnetic parameters for this star, but further and more detailed analysis through synthetic modeling is needed.\smallskip

\begin{equation}
    B_d = 20 \times B_1^{max}\frac{3-u}{15+u} (\cos\beta \cos i + \sin \beta \sin i)
    \label{eq:mod_theoretical}
\end{equation}

\begin{table}[ht!]
\caption{\label{tab:i}Estimation of the radius and inclination angle and magnetic angle of IRAS~17449+2320 for two different effective temperatures.}
\centering
\begin{tabular}{ccccc}
 \hline\hline
    T$_{eff}$ (K) & R (R$_\odot$) & i ($^{\circ}$) & $\beta$ ($^{\circ}$) & B$_d$\\
    \hline
    10\,000 & 2.8 & 50 & 79 & 8\,180\\
    12\,000 & 1.9 & 66 & 86 & 5\,275\\         
 \hline                                  
\end{tabular}
\end{table}

The longitudinal magnetic field is well fit by a first-order sinusoid, pointing to a mainly dipolar configuration. However, the magnetic field modulus shows that there could be a quadrupolar contribution in addition to it, as discussed in sec \ref{sec:b}. If we compare Fig. \ref{fig:bz-ls} and \ref{fig:mag-mod}, it can be seen that there is an offset of the maximum and minimum values of the longitudinal magnetic field and magnetic field modulus. This offset could indicate a departure from the axial symmetry in the magnetic field \citep{Preston1970}. All the characteristics of the magnetic field observed until now are compatible with the Oblique Rotator Model and with a rigid rotator. Detailed modeling of the magnetic field will be based on a more extensive dataset currently being acquired.

\subsection{Origin of the equivalent width variation}\label{subsec:EW}
As mentioned in Section \ref{sec:ew}, there are some metallic lines that present equivalent width variations with time. This equivalent width variation can be produced by inhomogeneities in the abundances of elements on the surface of the star, such as chemical spots, although more detailed modeling is needed to confirm it (Bermejo Lozano et al., in prep.). Chemical spots would be interesting but exceptionally unusual for IRAS~17449+2320. Chemical spots are commonly found in strongly magnetic Ap and Bp stars. These spots are associated with strong over-abundances of many elements in the stellar atmospheres. This is widely believed to be generated by atomic diffusion under the influence of a magnetic field \citep{Michaud1981}. IRAS~17449+2320 is not classified as a Bp star, and it does not display the strong absorption lines characteristic of Ap/Bp stars, Am stars, or HgMn stars. Thus, if the equivalent width variations are due to chemical spots, this would represent a new class of chemically spotted star. However, this EW variation could also be another manifestation of the Zeeman effect, called magnetic intensification \citep{Stift2003}. This magnetic intensification is a desaturation of the line due to unresolved Zeeman splitting, and as such, it only appears in stronger lines that are partially saturated.\smallskip

The equivalent width variation follows the same tendency as the magnetic field modulus, showing deeper lines at the minimum of $|B|$ (Figure \ref{fig:period-test}). The EW variation is expected to follow the absolute value of the longitudinal magnetic field or the magnetic field modulus. Finding the minimum of the EW at the negative extremum of B$_z$ is compatible with the magnetic intensification hypothesis. However, in our case, we are seeing this minimum of EW at the positive extremum, which contradicts this hypothesis.\smallskip

However, magnetic intensification is not the only possible explanation for the equivalent width variation. Variable veiling could cause EW variability. Veiling is the "filling-in" of stellar absorption lines by extra continuum emission, making these lines appear weaker \citep{Hartigan1991}, and it is often assumed to be caused by an accretion shock. Such extra continuum emission might explain the ultraviolet (UV) excess reported by \cite{Korcakova2022}. They compared the UV emission of 2~Cet, a Be type star with an effective temperature of $\sim$ 9\,800 K, with the UV emission of IRAS~17449+2320. For the NUV filter, they obtained an absolute brightness of 3.9 mag for 2 Cet and between 2.4 - 2.7 mag for IRAS~17449+2320.\smallskip

Veiling could vary with the rotation period of the star, if there is an accretion shock at or very near the surface of the star, and it is rotating in and out of view. However, it is unlikely that this would be stable over a time span of years, since the veiling would also vary with the intrinsic strength of the accretion shock, and hence on the momentary accretion rate and velocity. This would lead to a strong component of variability that is independent of rotation, as observed in T-Tauri stars \citep{Hartigan1991}.

Another possibility is that we are seeing a variable line emission partially filling in absorption lines. If the gas generating line emission was trapped by the magnetic field of the star, then the emission could plausibly be modulated with the rotation period. However, it is not expected that this phenomenon would be stable for a decade or more. On a careful inspection of these lines, we see no evidence for a circumstellar emission or absorption component, and no evidence for variability outside the photospheric line. But the \ion{Fe}{II} 4515~\AA~line looks fairly weak in Fig \ref{fig:spectrum+vstokes}, with no real sign of emission. Therefore, it is unlikely that line emission is causing this EW variability. Detailed spectral modeling, including the Zeeman effect and detailed chemical abundances, could help resolve this ambiguity (work in progress).

\subsection{Binarity}\label{subsec:binarity}
Binarity has been the only scenario suggested to explain not only IRAS~17449+2320, but all FS~CMa stars \citep{Miroshnichenko2007, Mennickent2017, Khokhlov2018} to date. \cite{IRASProceedings2025} suggests that IRAS~17449+2320 is a contact binary or an ex-triplet system, the magnetic star is the result of a merger, and its companion is a long-distance sdO- or sdB-type star. However, there are no spectroscopic features in the spectra of IRAS~17449+2320 that point to the presence of a secondary object. In Fig. \ref{fig:spectrum+vstokes}, we conclusively demonstrated that this splitting in the lines is produced by the magnetic field, each component of the split corresponds to the phases of the circular polarization. In addition, the circular polarization is observed along the whole spectral range. The reason why only certain lines present a splitting is related to their wavelength and Landé factor of the line, i.e., lines with a higher wavelength and Landé factor will show a clearer splitting of their sigma components, rather than with the presence of a cool spectroscopic companion. Also, some of these metallic lines appeared as triplets, i.e., \ion{O}{I} 7772, 7774, 7775 \AA, which discards the possibility that these lines are produced by a spectroscopic companion. \smallskip

Moreover, the radial velocity obtained from the LSD $I$ profiles does not show traces of binarity (Fig. \ref{fig:rv}). The variation of the radial velocity is small, around 2~km~s$^{-1}$ between the highest and lowest value, and does not show any periodic tendency. The distribution of the measurements points to a mean value of $\sim$-17.55 $\pm$ 0.11~km~s$^{-1}$. In addition, we performed a time series analysis over the radial velocities calculated, using the Lomb-Scargle method. There are no significant frequencies detected for the radial velocity, as the false alarm probability is $\sim$1 for all of them.\smallskip

\cite{IRASProceedings2025} also determined a 36-day period based on the ratio of the minima in flux on the red and blue sides of H$_\alpha$. This variation of the intensity of the absorption of the wings could be produced from plasma confined by the magnetic field of the star, although there are other sources of emission contributing to H$_\alpha$. Other magnetic B stars also show emission in H$_\alpha$, but they have profiles with substantially different strengths and morphologies from this star. A more detailed analysis of the circumstellar envelope of IRAS~17449+2320 is needed in order to explain all the emission features of this object.

In conclusion, the arguments given to support the binarity scenario for IRAS~17449+2320 can be explained through the magnetic field of the star. And, in addition, there is no other clear evidence of binarity in this object with the current data available.

\subsection{Origin of IRAS~17449+2320}\label{subsec:origin}

Strong, stable, large-scale magnetic fields are observed in certain types of stars, such as Ap/Bp stars \citep{Donati2009}, a small population of OB stars \citep{Grunhut2017, Alecian2014}, and some sub-dwarfs \citep{Shulyak2019} and white dwarfs\citep{Ferrario2015}. Several hypotheses have been proposed to explain their origin, with the fossil field hypothesis being the most widely supported. This hypothesis maintains that these magnetic fields are stable remnants from an earlier stage of evolution.  However, the origin of the initial magnetic field that formed the fossil remains a topic of debate. One possibility is that the fossil field formed during the pre-main sequence phase of the star.  Intermediate-mass stars likely exhibit strong convection during the pre-main sequence stage and, therefore, likely generate dynamo magnetic fields early in the pre-main sequence. These magnetic fields may "freeze" during stellar evolution as a radiative envelope develops and remain in time if they are stable enough \citep{Mestel1999, Moss2001, BraithSprout2017}. However, this does not explain the low incidence of A and B stars with detectable magnetic fields (from $\sim$1 to $\sim$5 \%, depending on mass; \citealt{Sikora2019}. Another scenario considered to explain the fossil fields is the stellar merger hypothesis. Theoretical models suggest that magnetic fields can be formed during the merger of a binary system and remain stable in time \citep{Schneider2019}. However, fossil fields are not the only scenario considered to explain the origin of magnetic fields.\smallskip

IRAS17449+2302 is the first B[e] star with a magnetic field detected and, due to the peculiar nature of this object, it is still not clear what the origin of this star nor of its magnetic field. The large amount of circumstellar matter surrounding the star and its space velocity could suggest that this star is the result of a merger. It may therefore be that the magnetic field was created during this merging process. The large amount of gas and dust surrounding this object cannot be produced by a single B-type star in the main sequence, and it could be the main characteristic pointing to a merger origin for this star. This material was created and released during the merging and created a shell around it. The slow rotation of the star could also be a consequence of this process. Right after the merger, the star was rotating near critical velocity. Afterwards, the magnetic field interacted with the circumstellar material released during the merger, causing the star to slow down. This process is known as magnetic braking. The loss of angular momentum of the star is not only produced by the magnetic braking, but also by the mass-loss process \citep[Section 3.1.]{Schneider2020}. Altogether, the resulting object of the merging process would be a slow-rotating star. However, not every magnetic massive star detected is a slow rotator. For example, HR~5907 is a B-type star with a magnetic field of $\sim$16 kG and a rotation period of $\sim$0.5 days \citep{Grunhut2013}. $\sigma$~Ori~E is another example of a rapidly rotating magnetic star, with a period of $\sim$1.19 days \citep{Oksala2012} and a magnetic field of $\sim$10 kG \citep{Townsend2010}. \cite{Wang2022} suggest stars that are merger products rapidly lose angular momentum through strong magnetic fields, leading to slow rotation on the main sequence, while stars that accreted from a disk remain rapid rotators. Although, it is still unclear how strong this magnetic braking process is and if there are any other mechanisms involved in the momentum loss, for example, the effects of the magnetic field orientation or the presence of a rigidly rotating magnetosphere. In addition, the space velocity of IRAS~17449+2320 could be consistent with this star being ejected from the birth cluster due to the merging process \citep{Dvorakova2024}.

The theoretical simulations of \cite{Schneider2019} showed that the resulting object of a merger process preserved practically all the mass of the progenitors. These simulations were performed for a resulting object of 17 M$_\odot$, as they were meant to explain the nature of $\tau$ Sco, a magnetic ‘blue straggler’ star. On the other hand, IRAS~17449+2320 has a stellar mass of 2-3 M$_\odot$. Therefore, it would be of special interest to run the theoretical simulations with similar characteristics to this star, as its progenitors present a different inner structure and a different evolutionary path than for $\tau$ Sco, which could affect the merging process and the resulting object. In these theoretical simulations, \citet{Schneider2020} assumed a geometrically thin and optically thick Keplerian circumstellar disk. In contrast, IRAS~17449+2320 shows a much more complex circumstellar envelope, based on its spectroscopic features, than was predicted by these theoretical simulations. This could point to a nonconservative mass transfer during the merger. This circumstellar setup would be harder to model, but the results will be interesting to compare with.\smallskip

\subsection{The role of IRAS~17449+2320 in the FS~CMa group}\label{subsec:role}
The first classification of the stars presenting the B[e] phenomenon was performed by \cite{Lamers1998}. They found that different types of objects were presenting the B[e] phenomenon, including compact planetary nebulae, symbiotic stars, Herbig Ae/Be stars, and B[e] supergiants. However, there were around 50 objects that could not be included in any of the previous categories, and they received the name of FS~CMa stars by \cite{Miroshnichenko2007}. The FS~CMa stars are a heterogeneous group, of which certain members are being reclassified nowadays, for example, CI~Cam \citep{Sidoli2022}.\smallskip

This missclasification introduces some uncertainty toward, IRAS~17449+2320. This star does not belong to other subgroups of objects presenting the B[e] phenomenon: it is far from a star formation region, so it is not a Herbig Ae/Be star \citep{Hillenbrand1992}. It is not a symbiotic star and not a supergiant, as it has no signs of a companion \citep{Allen1983}, and its mass is 2-3~M$_\odot$ \citep{Bouret2012}. It is not in a compact nebulae as there are no high ionization lines, such as [\ion{O}{II}]. But it also presents significant differences with other members of the FS~CMa group.\smallskip

The photospheric spectrum of IRAS~17449+2320 is much clearer, i.e., the absorption lines present less contamination from the circumstellar medium than other stars, such as HD~50138 or MWC~342. It also present fewer forbidden lines than other objects, showing only [\ion{O}{I}]~6630 and 6363~\AA. And it is the only star from this group with resolved Zeeman split components up to now.\smallskip

Therefore, is IRAS~17449+2320 a unique object among FS~CMa stars? Studying more members of this group is of special interest, as FS~CMa stars could be an overlooked channel for intermediate-mass mergers, and they could provide a better understanding of the merger process.

\section{Conclusions}
IRAS~17449+2320 is the first B[e] star with a detected magnetic field ($\sim$6\,000 G). Its magnetic field was first identified through the Zeeman splitting of some lines in its spectrum, such as \ion{N}{I}, \ion{Si}{II}, and \ion{Ca}{II} \citep{Korcakova2022}. In order to study this magnetic field in more detail, we analyzed spectropolarimetric observations. Through the Stokes $V$ parameter of these observations, we measured the longitudinal magnetic field of the star. The longitudinal magnetic field presents a periodic sinusoidal variation, with an estimated rotational period of 36.11 $\pm$ 0.01 days.\smallskip  

The sinusoidal variation of B$_z$ suggests a mainly dipolar configuration of the magnetic field. However, the magnetic field modulus hints at a possible quadrupolar component, although further observations are needed in order to verify this possibility.\smallskip

Zeeman splitting is not the only effect that can be observed in the spectrum of this star. Some metallic lines, such as \ion{Ti}{I}, \ion{Mg}{II}, and \ion{Fe}{II}, present a variation in their equivalent width with the same periodicity as the longitudinal magnetic field. These lines could show magnetic intensification.\smallskip

Previous studies suggested that the splitting observed in multiple lines of the spectrum of this star was evidence of the composite spectrum of a binary system. After performing the analysis, we conclude that the splitting is produced by the Zeeman effect, and we did not observe evidence of binarity. The magnetic field, together with the large amount of circumstellar matter, the space velocity, and the lack of clear evidence of a companion, suggests that IRAS~17449+2320 could be the result of a merger. Theoretical models \citep{Schneider2019} show that magnetic fields can be created during a merger process and remain stable on evolutionary timescales.\smallskip

The determination of the rotational period of IRAS 17449+2320 provides important groundwork for future studies of the star. Such studies could include detailed modeling of the spectrum to determine temperature, $\log g$, and chemical abundances. However, such modeling must include the Zeeman effect to produce reliable results.  With more spectropolarimetric observations, providing denser rotation phase coverage, the star may also be a candidate for Zeeman Doppler Imaging to provide a surface magnetic field map.\smallskip

FS CMa stars, and particularly IRAS~17449+2329, offer a great possibility to study processes related to the merger process, such as ejection and re-accretion of matter, creation of the magnetic field, and changes in the chemical composition of the star and in the internal structure of young post-merger objects, which could present variations in the pulsation modes of the stars. The detailed description of all these physical processes is important for our understanding of star formation. The merger process may play a crucial role there, especially in massive stars that cannot be created straightforwardly by the collapse of a molecular cloud. Since FS~CMa stars are isolated objects, all these processes are easier to study than in objects in star-forming regions that are crowded by stars and full of gas and dust.

\section*{Data Availability}
The mask used to obtain the LSD profiles is uploaded at the Zenodo repository: \href{https://zenodo.org/records/19915806}{https://zenodo.org/records/19915806}. Tables \ref{tab:b} and \ref{tab:ew} are available in electronic form at the CDS: \href{http://cdsweb.u-strasbg.fr/cgi-bin/qcat?J/A+A/}{http://cdsweb.u-strasbg.fr/cgi-bin/qcat?J/A+A/}. The spectra used in this work are available from the CFHT archive at the Canadian Astronomy Data Centre (CADC): \href{https://www.cadc-ccda.hia-iha.nrc-cnrc.gc.ca/en/cfht/}{https://www.cadc-ccda.hia-iha.nrc-cnrc.gc.ca/en/cfht/}.

\begin{acknowledgements}
    IBL acknowledges the Charles University Grant Agency (project number 6124). GAW acknowledges Discovery Grant support from the Natural Sciences and Engineering Research Council (NSERC) of Canada. CPF acknowledges funding from the European Union's Horizon Europe research and innovation programme under grant agreement No. 101079231 (EXOHOST) and from the United Kingdom Research and Innovation (UKRI) Horizon Europe Guarantee Scheme (grant number 10051045).
    
    This work is based on observations obtained with ESPaDOnS at the Canada-France-Hawaii Telescope (CFHT), which is operated by the National Research Council (NRC) of Canada, the Institut National des Science de l'Univers of the Centre National de la Recherche Scientifique (CNRS) of France, and the University of Hawaii. The observations at the Canada-France-Hawaii Telescope were performed with care and respect from the summit of Maunakea which is a significant cultural and historic site. It has also made use of the VALD database, operated at Uppsala University, the Institute of Astronomy RAS in Moscow, and the University of Vienna.

\end{acknowledgements}

\bibliographystyle{aa} 
\bibliography{thebibliography} 

\begin{appendix}
\section{Stoke $Q$ and $U$ profiles}\label{app:1}
The two phases of Stokes $Q$ and $U$ observations show linear polarization in LSD profiles, with a strong variation. No linear polarization is confidently detected in individual lines. Stokes $Q$ and $U$ profiles are sensitive to the transverse component of the stellar magnetic field \citep{Wade2000-linearpol}. Both profiles present a positive central lobe, together with two negative lobes on the sides. The amplitude of the $Q$ and $U$ profiles is substantially lower than that of the Stokes $V$ profiles (by a factor of more than three compared to the Stokes $V$ profile with the highest amplitude). Thus the observed Stokes $V$ profiles are unlikely to be affected by any instrumental crosstalk from linear into circular polarization, since this crosstalk for ESPaDOnS is typically less than 1\%.

\begin{figure}[ht]
    \centering
    \includegraphics[width=0.5\textwidth]{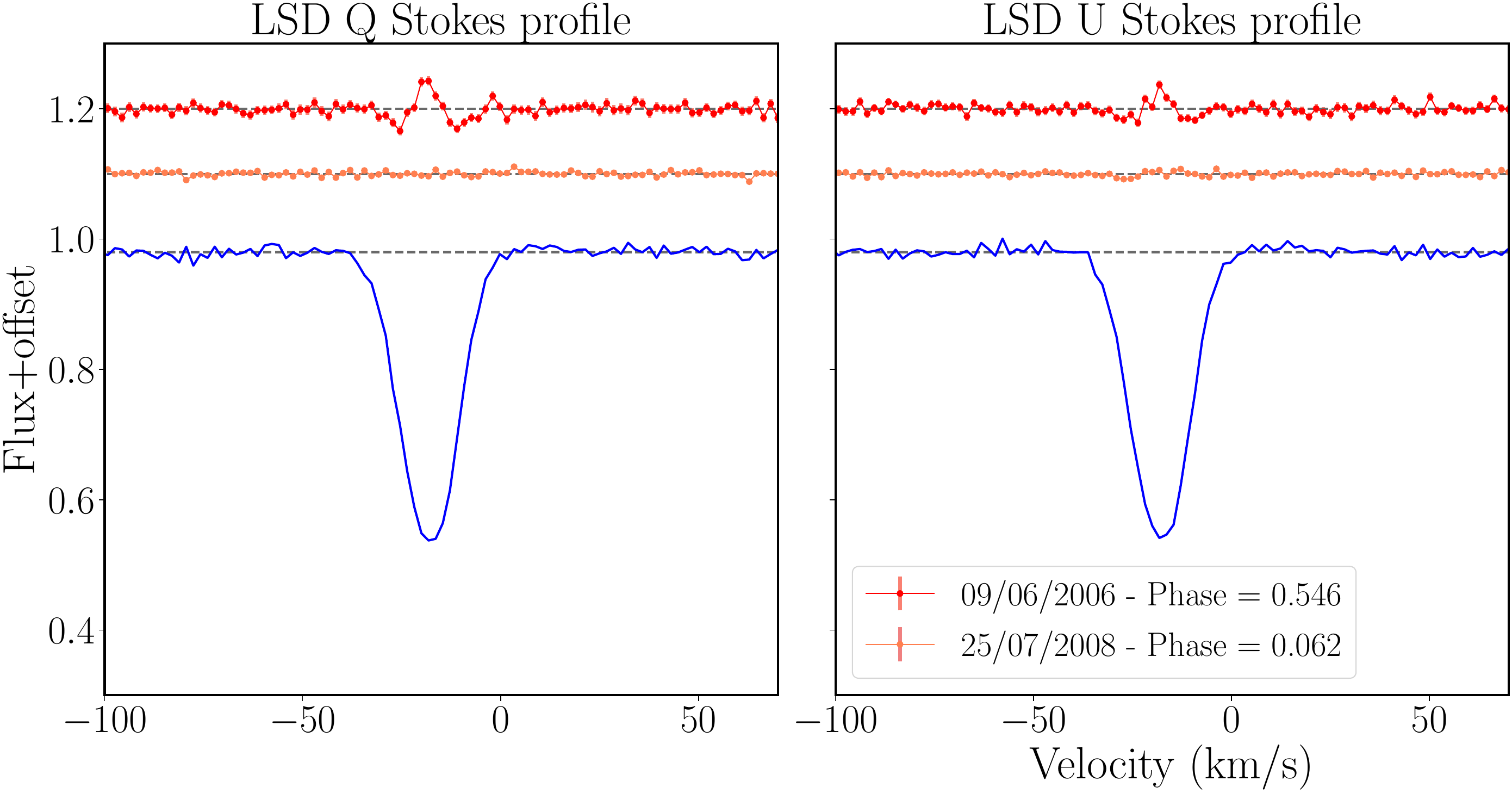}
    \caption{LSD Stokes $Q$ and LSD Stokes $U$ profiles from the nights of 9 June 2006 (upper profiles) and 25 July 2008 (lower profiles) showing linear polarization.}
    \label{fig:QU}
\end{figure}

\section{Polarization of emission and resonance lines}

IRAS 17449+2320 presents multiple lines that show some emission contribution (absorption and emission profile), as shown in Fig. \ref{fig:spectrum+vstokes}. Here, we show the polarization of two resonance line doublets, \ion{Na}{I} D1, D2, and \ion{Ca}{II} H, K. Both doublets show broad emission overlapped with the absorption component, and do not present circular nor linear individual polarization. The forbidden [\ion{O}{I}] 6300, 6363 \AA\ do not show any polarization either. These lines provide some important information about the circumstellar matter surrounding the central star, although further analysis is needed.

\begin{figure*}[ht]
    \centering
    \includegraphics[width=\textwidth]{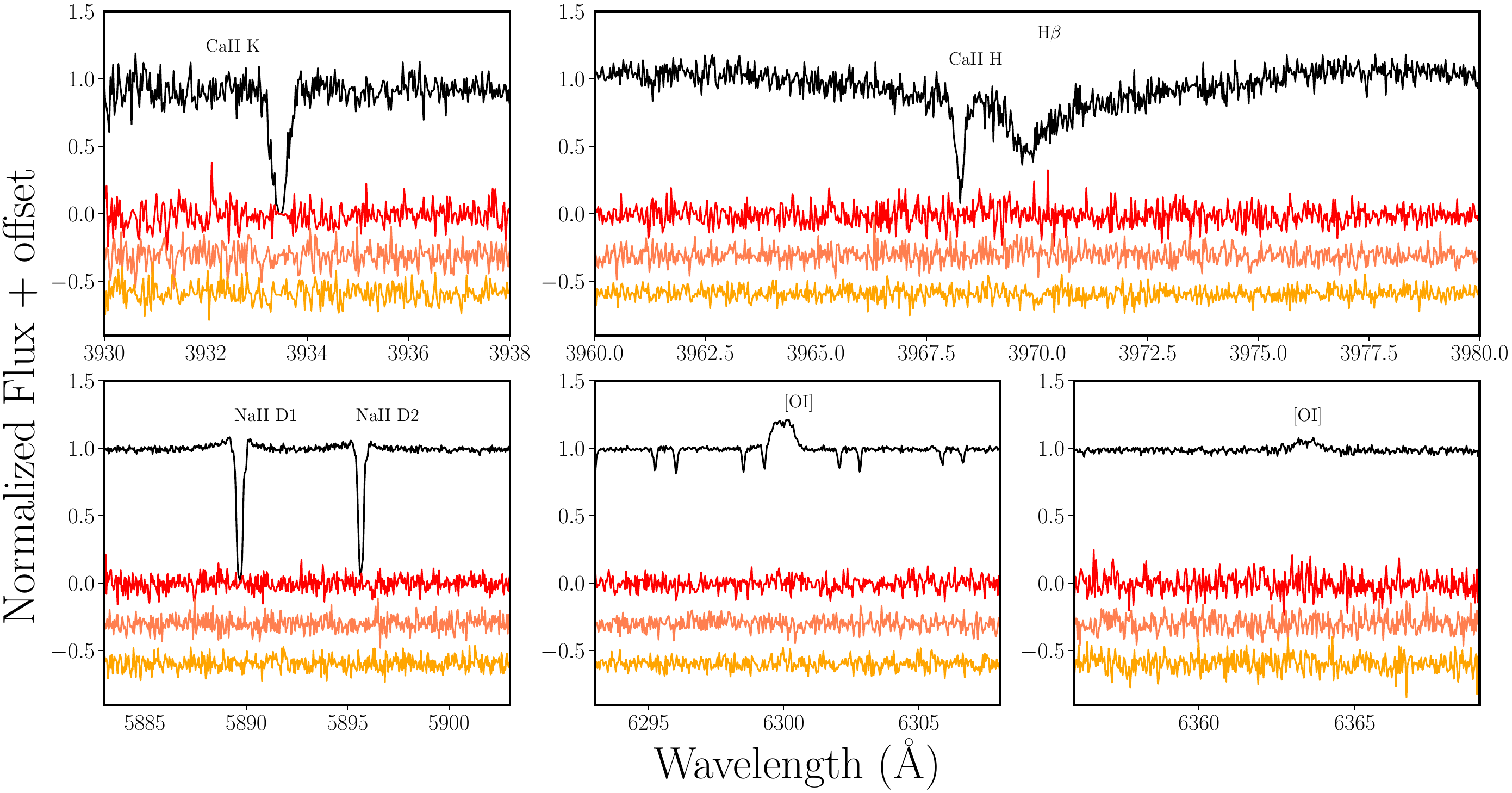}
    \caption{Spectrum (black), circular polarization, Stokes $V/I_c$ (red), and linear polarization, Stokes $Q/I_c$ and Stokes $U/I_c$ (coral and orange, respectively) of the forbidden emission lines, [\ion{O}{I}]~6300, 6363~\AA, and the resonance doublets \ion{Na}{I} D1, D2 and \ion{Ca}{II} H, K of IRAS~17449+2320 from the night of 9 June 2006. The Stokes $Q$ and Stokes $U$ are shifted -0.3 and -0.6, respectively, for visualization purposes. In the three panels of the lower row, Stokes $V$, $Q$ and $U$ are multiplied by four for a better visualization.}
    \label{fig:spectrum+vstokes_emission}
\end{figure*}

\section{Zeeman splitting patterns}\label{app:quantum}

\begin{figure}[ht]
    \centering
    \includegraphics[width=\hsize]{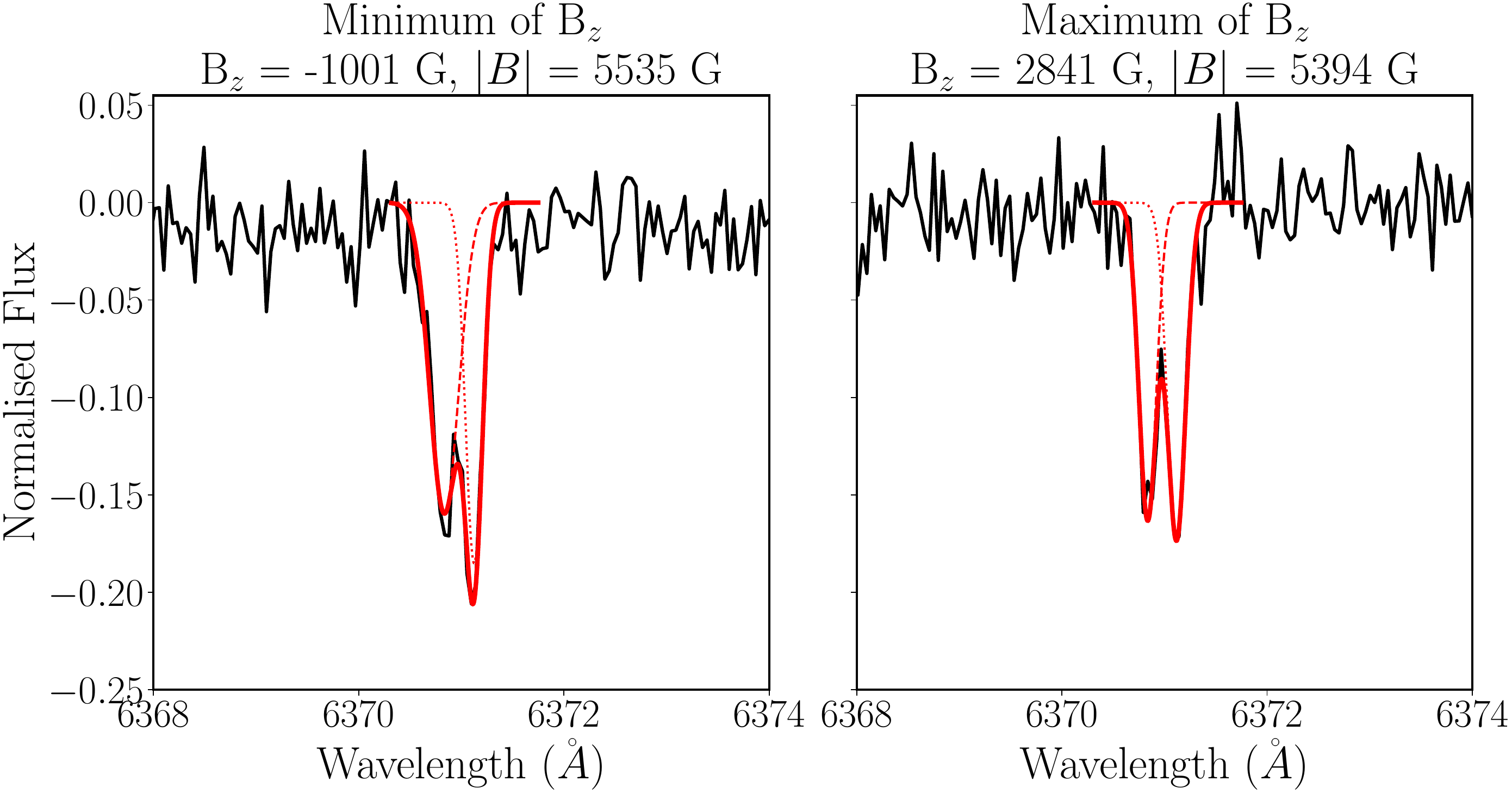}
    \includegraphics[width=\hsize]{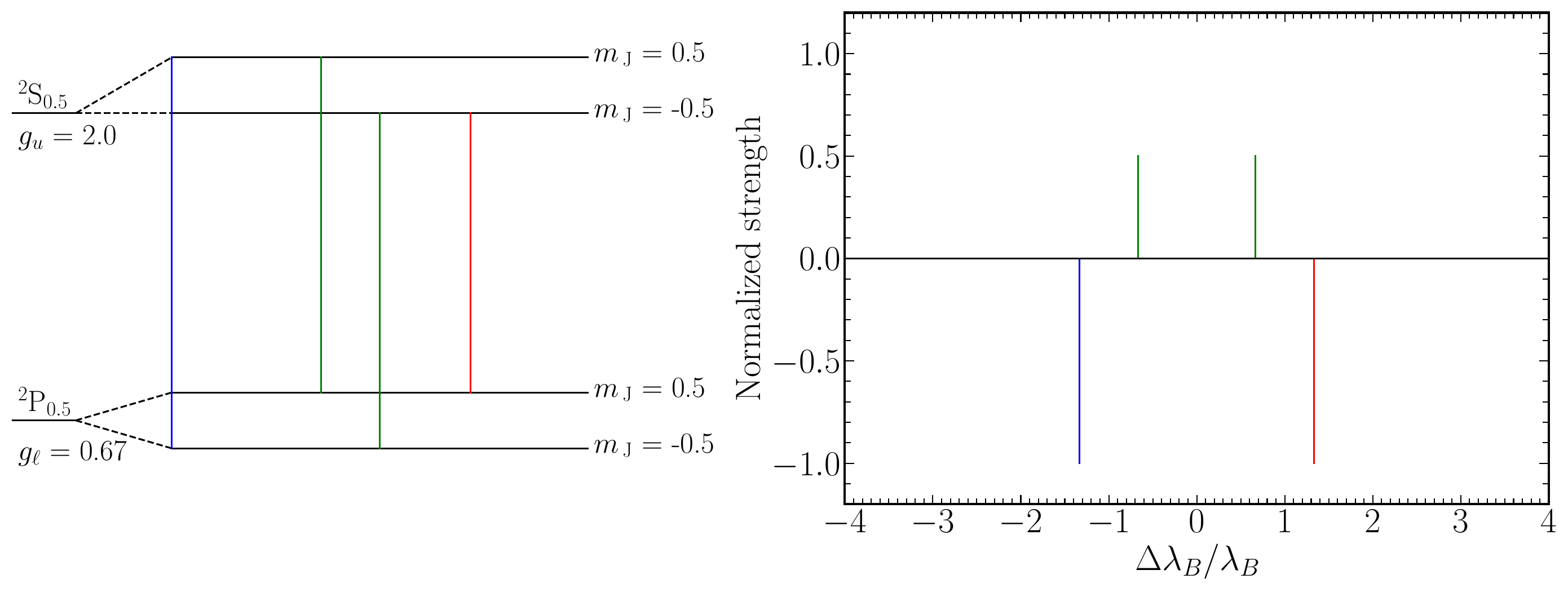}
    \caption{Gaussian fit to the two Zeeman split components of \ion{Si}{II} 6371\AA, as in Fig. \ref{fig:zeeman}. This \ion{Si}{II} line corresponds to the transition of $^{2}S_{1/2}$ -- $^{2}P_{1/2}$.}
    \label{fig:zeeman_si}
\end{figure}

\begin{figure}[ht]
    \centering
    \includegraphics[width=\hsize]{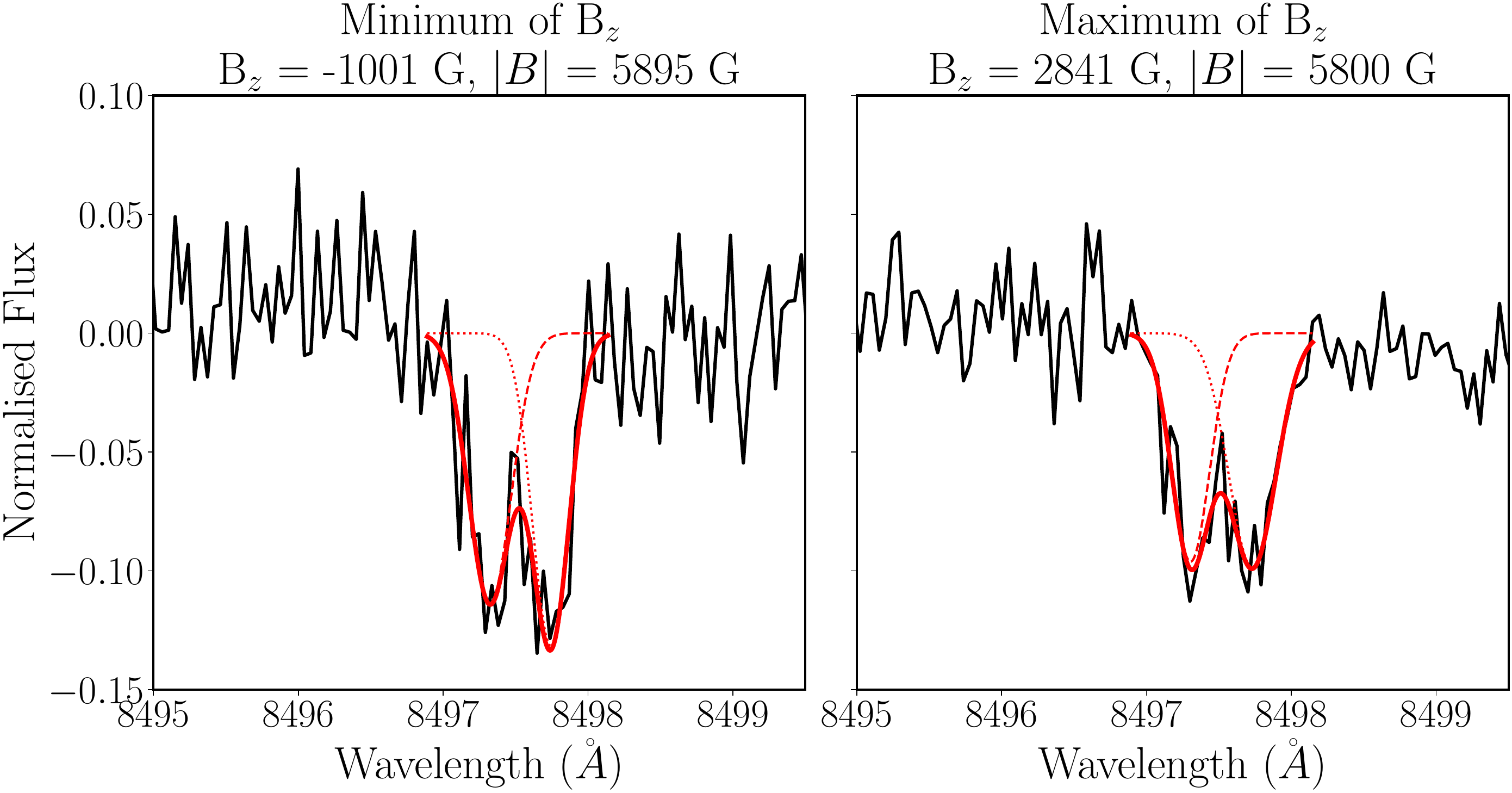}
    \includegraphics[width=\hsize]{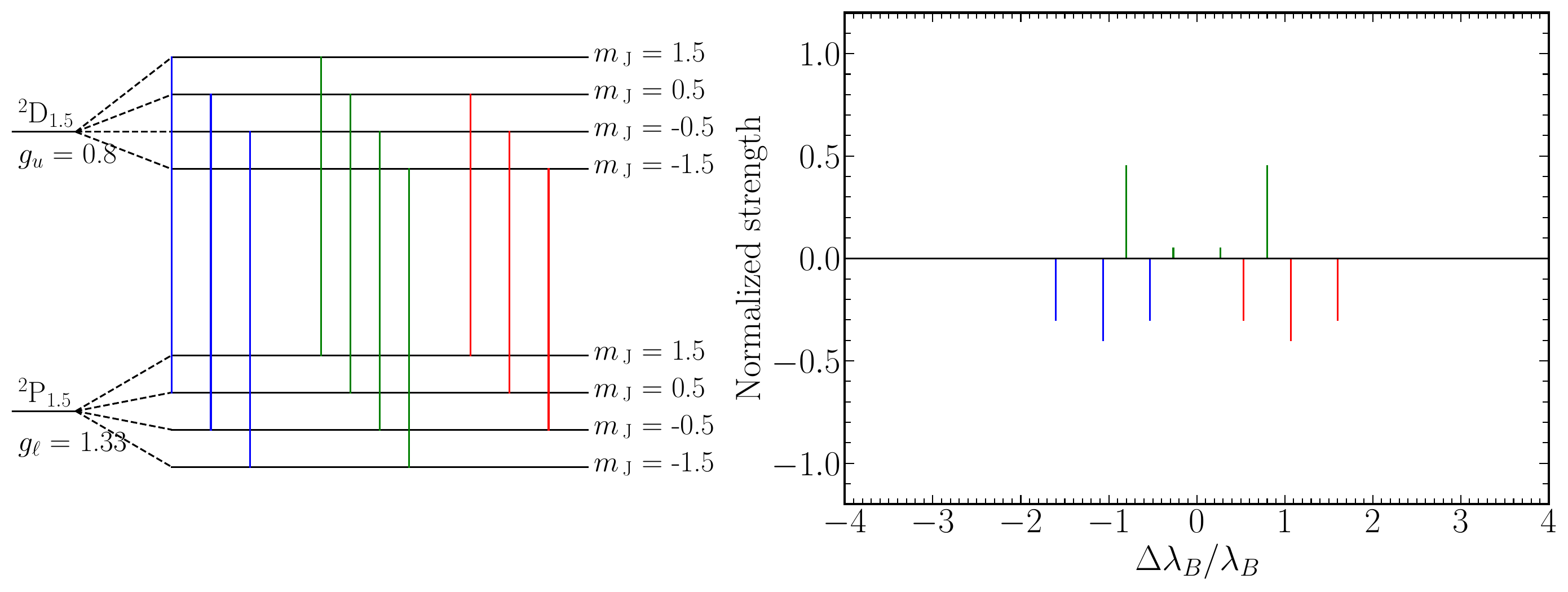}
    \caption{Gaussian fit to the two Zeeman split components of \ion{Ca}{II} 8497\AA, as in Fig. \ref{fig:zeeman}. This \ion{Ca}{II} line corresponds to the transition of $^{2}D_{3/2}$ -- $^{2}P_{3/2}$.}
    \label{fig:zeeman_ca}
\end{figure}

The magnetic field modulus is obtained through three different lines in this paper, as explained in Section \ref{sec:b}. These three different lines correspond to different transitions. In this appendix, we show more detail for the \ion{Si}{II} and \ion{Ca}{II} lines. The \ion{Si}{II} transition is similar to the \ion{N}{I} transition shown in Fig. \ref{fig:zeeman}, as can be seen in Fig. \ref{fig:zeeman_si}. However, the \ion{Ca}{II} transition shows a more complex Zeeman pattern (Fig. \ref{fig:zeeman_ca}). This complexity can also be found in the splitting of the line, as there are extra components in the core of the two main groupings.

\end{appendix}

\end{document}